\documentclass[prb,floatfix,amsmath,twocolumn,notitlepage,superscriptaddress]{revtex4-2}

\usepackage{tikz} 
\usepackage{amsmath,amsthm,amsfonts}
\usepackage{mathtools}
\usepackage{verbatim}
\usepackage{hyperref}
\usepackage{dsfont}
\usepackage{soul} 
\usepackage{array, makecell}
\usepackage{tabularx}
\usepackage{enumitem}
\usepackage{calc}

\theoremstyle{definition}

\newcommand{\ket}[1]{|#1\rangle}

\definecolor{cool_green}{rgb}{0.0, 0.5, 0.0}
\definecolor{new_pink}{rgb}{1, 0.0, 1}

\begin{document}

\title{Non-zero Momentum Implies Long-Range Entanglement When Translation Symmetry is Broken in 1D}

\author{Amanda Gatto Lamas}\email{amandag6@illinois.edu}

\author{Taylor L. Hughes}\email{hughest@illinois.edu}
\affiliation{Department of Physics and Anthony J. Leggett Institute for Condensed Matter Theory, University of Illinois at Urbana-Champaign,  Urbana,  IL 61801, USA}

	\date{\today}
	
\begin{abstract}

A result by Gioia and Wang [Phys Rev X 12, 031007 (2022)] showed that translationally symmetric states having nonzero momentum are necessarily long range entangled (LRE).  Here, we consider the question: can a notion of momentum for non-translation symmetric states directly encode the nature of their entanglement, as it does for translation symmetric states? We show the answer is affirmative for 1D systems, while higher dimensional extensions and topologically ordered systems require further work. While Gioia and Wang’s result applies to states connected via finite depth quantum circuits to a translation symmetric state, it is often impractical to find such a circuit to determine the nature of the entanglement of states that break translation symmetry. Here, instead of translation eigenstates, we focus on the many-body momentum distribution and the expectation value of the translation operator in many-body states of systems having broken translation symmetry. We show that in the continuum limit the magnitude of the expectation value of the translation operator $|\langle T \rangle|$ necessarily goes to $1$ for delocalized states, a proxy for LRE states in 1D systems. This result can be seen as a momentum-space version of Resta's formula for the localization length. We investigate how accurate our results are in different lattice models with and without well-defined continuum limits. To that end, we introduce two models: a deterministic version of the random dimer model, illustrating the role of the thermodynamic and continuum limits for our result at a lattice level, and a simplified version of the Aubry-Andre model, with commensurate hopping for both momentum and position space. Finally, we use the random dimer model as a test case for the accuracy of $|\langle T \rangle|$ as a localization (and thus entanglement) probe for 1D periodic lattice models without a well-defined continuum limit.

\end{abstract}

\maketitle

\section{Introduction}

In Ref. \cite{nonzero_momentum}, Gioia and Wang showed that a translation-symmetric state having a nontrivial momentum, i.e., a nontrivial expectation value of the lattice translation operator $\langle T\rangle = \langle e^{ia\hat{p}}\rangle \neq 1$, implies that the state is long-range entangled (LRE). This result is valid for any states that are adiabatically connected, or, in quantum circuit language, connected via a finite-depth circuit, to a state having nontrivial momentum. Thus, implicitly, even states that break translation symmetry can be identified as LRE provided that there exists a finite depth quantum circuit connecting them to translation-symmetric states having nontrivial momentum. This motivates the questions: can a notion of momentum for non-translation symmetric states directly encode the nature of their entanglement? Moreover, how does the expectation value of the translation operator encode LRE in generic states that break translation symmetry?

Here we seek to answer these questions. Unfortunately, an approach based on finding a circuit to transform a generic translation-broken state to a translation invariant state is highly nontrivial. Therefore, it is desired to find a measure to determine the entanglement characterization of a generic state that depends on only the translation properties of a state itself, and not on its connection to another translation-symmetric state. For this measure we choose the  expectation value of the translation operator $z_t = |\langle T \rangle|.$ For an explicit context we focus on one-dimensional systems where the spatial localization properties of a state are a good proxy for its entanglement properties. Indeed, localized states in 1D systems are short-range entangled, while delocalized states are long-range entangled. In this context we show that $z_t$ captures the spread of the total momentum distribution, revealing localization properties of the state, and by proxy its entanglement properties. 

 To accomplish this, we establish a dual picture for Resta's formula for the localization length. Resta's formula is based on the expectation value of the modular position operator $z_x\equiv \vert\langle U\rangle\vert =\vert\langle e^{i\frac{2\pi}{L}\hat{X}}\rangle\vert$,  where $\hat{X}$ is the many-body position operator {\cite{Resta-98-operator,Resta-99, 2008-Massar-Spindel,Aharonov:1969qx}}. The localization length can be determined via\begin{equation}
  z_x= e^{-\frac{1}{2}\left(\frac{2\pi}{L}\right)^2\lambda_x^2},\label{eq: Resta's formula}
\end{equation}
where $L$ is the system size, and the localization length $\lambda_x$ is defined as the square root of the second cumulant (spread) of the total position distribution, $\lambda_x = \sqrt{C^{(X)}_2}$. For localized systems, $z_x$ tends to unity as system size increases. By obtaining an analogous expression for localization in momentum space, we show that $z_t$ tends to unity for delocalized states having broken translation symmetry. For translation symmetric states, $z_t$ is exactly unity whether the state is localized or delocalized. 

In addition to studying this measure of LRE, a major goal of our work is to sharpen the role of momentum-space-based measures of localization and entanglement for systems that break translation symmetry. Several works have used momentum-space-based quantities to study localization and entanglement \cite{2008-Massar-Spindel, clock_shift_2017,Concurrence-disordered-2008,momentum-space-shannon-entropy-2012,taylor-dimer-momentum-spectrum}, including in systems that break translation symmetry. In particular, Refs. \cite{2008-Massar-Spindel, clock_shift_2017} used the uncertainty between the translation operator $T$ and the modular position, or twist, operator $U$ to study systems having localization transitions resulting from disorder or slowly varying potentials. However, their focus was not on the momentum distribution or entanglement properties of the states as we consider here. Additionally, other works such as \cite{Concurrence-disordered-2008, taylor-dimer-momentum-spectrum,momentum-space-shannon-entropy-2012} used entanglement measures in momentum space to study localization transitions in disordered and quasiperiodic systems, but an explicit relationship with the translation operator and the momentum distribution was absent. More recently, Ref. \cite{chen_theory_2025} developed a framework to study localization of quasiperiodic systems that bridges the momentum and position space localization pictures, but their framework has not yet been extended to generic systems that break translation symmetry. In the context of these previous works and Ref. \cite{nonzero_momentum}, our goal is to explicitly formalize the connections between the expectation value of the translation operator, the many-body momentum distribution, and the localization/entanglement properties of 1D systems.

This article is organized as follows: in Sec. \ref{sec: background} we provide a review of Ref. \cite{nonzero_momentum} and of uncertainty relations for unitary operators, which place our main result in context. In Sec \ref{sec: prob distr and char funcs} we review the background of probability distributions and characteristic functions, and use those tools to derive our main result. In Sec. \ref{sec: numeric examples} we use our main result to determine the localization properties of different lattice models, numerically validating our claims.  
While our main focus is on the magnitude $z_t$, Ref. \cite{nonzero_momentum}, determined that the phase of the expectation value of $T$ for a LRE state will be sensitive to twisting boundary conditions/flux insertion. Hence, in Sec. \ref{sec: flux insertion} we discuss whether the phase $\arg\langle T\rangle$ is still sensitive to flux when translation symmetry is broken, and in what sense this flux sensitivity may still be used to distinguish between LRE and SRE states. Finally, we summarize our findings and point to future directions in Sec. \ref{sec: conclusion}.

\section{Background and overview of main results} \label{sec: background}

\begin{figure}
     \centering
        \includegraphics[width=\linewidth]{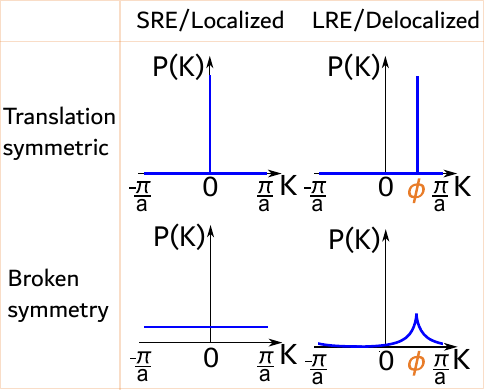}
         \caption{Illustration of sample momentum distributions for 1D systems having twisted periodic boundary conditions with twist $\phi$. Translation-symmetric states (first row) have delta-function-like momentum distributions. For SRE/localized states the peak is always centered at trivial momentum $K_0=0$ (or $\pi$, for fermions) even after twisting boundary conditions. For LRE/delocalized states, even a state that starts at a trivial value $K_0(\phi=0)=0$ will have its momentum shifted to $K_0(\phi)=\phi$ when subject to twisted boundary conditions. We can compare the distributions of the top row of translation symmetric states to the distributions for the translation-broken states in the bottom row.  Localized states that break translation symmetry have uniform, or flat, momentum distributions, and delocalized states may have single peaks with finite-width or distributions of peaks (see Section \ref{sec: AA model}). Upon twisting boundary conditions, the LRE/delocalized states have their momentum-distribution peaks shifted, and can be distinguished from the un-shifted distributions. In contrast, for SRE/localized states the distributions are shifted by twisting boundary conditions, but are not distinguishable from the un-shifted cases, and the many-body momentum is unchanged.}
        
         \label{fig: distribution schematic}
\end{figure}

While our goal is to understand the relationship between the entanglement properties and the momentum distribution of states that break translation symmetry, it is instructive to first review the results of Ref. \cite{nonzero_momentum} for translation-symmetric states. Consider a translation invariant many-body ground state that is an eigenstate of momentum. Since this state has a definite value of the many-body momentum $K_0$, the many-body momentum distribution is a delta function peaked at $K_0$. 

Such a translation-symmetric state could be long-range or short-range entangled (SRE), and a distinguishing characteristic is the value of the many-body momentum $K_0$, and how this value changes (or not) when the boundary conditions are twisted via flux insertion \cite{nonzero_momentum}. Indeed, in Ref. \cite{nonzero_momentum}, the authors prove that SRE states, which correspond to localized states for one-dimensional periodic systems, necessarily have trivial total momentum $K_0=0$ ($K_0=0,\pi/a$ for fermions). Moreover, the momentum of SRE states does not change when boundary conditions are twisted/flux is inserted. Hence, a localized, translation-symmetric state has a momentum distribution that is a delta-function peaked at a trivial value, no matter what flux is inserted in the system. 

In contrast, long range entangled states may have any value of momentum $K=\phi$, including, but not limited to, the trivial ones. Consequently, the state is delocalized if the momentum distribution is either peaked at a nontrivial value of momentum, or if it can be tuned to a nontrivial value of momentum via flux insertion. This is illustrated in the first row of Fig. \ref{fig: distribution schematic}, where the momentum distribution for translation invariant states is depicted. The two columns represent states that are localized or delocalized, respectively. While both cases represented in the figure are subject to a boundary condition twist, only the LRE state exhibits a flux sensitivity of the peak of the distribution. Therefore, the total momentum eigenvalue, obtained via the phase of the expectation value of the translation operator $\arg \langle \hat{T}\rangle$, can be used to distinguish between SRE (localized) and LRE (delocalized) translationally-invariant states in one-dimensional periodic systems. 

This simple picture applies for translation invariant states. However,  our goal is to study the entanglement properties of states that \emph{break} translation symmetry. States that are not translation symmetric are not eigenstates of the translation operator, and consequently do not have a definite value of many-body momentum. In fact, the many-body momentum distribution $P(K)$  (to be precisely defined in the next section), is generically no longer a delta function. Despite this change, we will demonstrate that the structure of this distribution is crucial for determining the entanglement characterization of a state. In fact, one of our main results is that the spread of this distribution (i.e., the second cumulant $C^{(K)}_2$) is captured by the \emph{magnitude} of the expectation value of the translation operator, $z_t\equiv|\langle {T}\rangle|.$ This is reminiscent of the well-known result that the position space cumulant is captured by the magnitude of the expectation value of the modular position operator.\cite{Resta-99,Souza_2000}

In fact, we will show that the expectation value of the translation operator is complementary to the localization properties determined from the modular position operator\cite{Resta-99,Souza_2000}. For example, the spread of the momentum distribution in momentum space is linked to the localization of the position distribution in position space via the uncertainty relation between the modular position $U=e^{i\frac{2\pi}{L}\hat{X}}$ and modular momentum $T = e^{ia\hat{K}}$ operators, where the modular momentum operator is the familiar translation operator. This relationship, which we review next, allows us to use $z_t$ as a probe for the localization and entanglement of many-body systems even when translation symmetry is broken.

To illustrate the general structure, let us consider systems having periodic boundary conditions. A state $\hat{X}|\Psi\rangle$, where $\hat{X}$ is the many-body position operator, is not well-defined with these boundary conditions, because it takes a viable state $|\Psi\rangle$ outside the Hilbert space of periodic states \cite{Souza_2000}. Therefore, the uncertainty in position space $\Delta x = \sqrt{\langle \hat{X}^2\rangle -\langle \hat{X}\rangle^2} $, which appears in the uncertainty principle $\Delta x\Delta p\geq \hbar/2$, is also not defined. Hence, the appropriate uncertainty relation between the position and momentum distributions of periodic systems must be obtained from the operators that are well-defined for periodic systems. Those are exactly the modular position $U=e^{i\frac{2\pi}{L}\hat{X}}$ and translation $T = e^{ia\hat{K}}$ operators.

The core of the uncertainty relation is in the commutation relation between $U$ and $T$, which depends on the total charge operator $\hat{N}_e$ as
\begin{equation}
    UT = e^{-i2\pi\frac{a}{L}\hat{N}_e}TU.\label{eq: UT commutation}
\end{equation}
Since we always assume that the many-body states under consideration have definite charge (are $U(1)$-symmetric), we can replace the operator $\hat{N}_e$ with the total charge $N_e$. When the filling $a{N}_e/L$ is an integer, the $U$ and $T$ operators commute and there is no nontrivial uncertainty relation restricting the possible values of their uncertainties. In contrast, when $a{N}_e/L\notin \mathbb{Z}$, there is an uncertainty relation that requires one distribution to be highly peaked when the other is highly spread, as we describe next. We provide more details of the uncertainty relations between the modular position and translation operators and the arguments below in Appendix \ref{appx: uncertainty relation}.

The uncertainties in $U$ and $T$ are given by
\begin{align}
    \Delta U^2 &= \langle \Psi|U^\dagger U|\Psi\rangle - \langle \Psi|U^\dagger|\Psi\rangle\langle\Psi|U|\Psi\rangle\nonumber\\
    &= 1 - |\langle\Psi|U|\Psi\rangle|^2 \nonumber\\
    &= 1-z_x^2,\\
    \Delta T^2&=1-z_t^2,
\end{align}
where $|\Psi\rangle$ is a generic many-body state in the Hilbert space.
Therefore, the uncertainty relation between the modular position and translation operators, derived in Refs. \cite{2008-Massar-Spindel,PRA_uncertainty}, is also an inequality relating $z_x$ and $z_t$. It can be written as 
\begin{equation}
    (1+2A)(1+z_x^2z_t^2)-(1+A)^2(z_x^2+z_t^2)+A^2\geq 0,\label{eq: uncertainty relation 2 zx zt}
\end{equation}
where $A = \big|\tan \left(\frac{aN_e}{2L} 2\pi\right)\big|$, for system size $L$.

As mentioned in the introduction, $z_x$ is related to the localization length in position space via Resta's formula (which we review in the next section). We will show that the uncertainty relation in Eq. \ref{eq: uncertainty relation 2 zx zt} implies that, when $[U,T]\neq0$, states that have a very localized momentum distribution, i.e., a small momentum spread $C^{(K)}_2/(2\pi/a)\ll1,$ have $z_t\to1$ and are  delocalized in position space. For the same reason, a state that is localized in position space, having a sharply localized position distribution captured by $z_x\to1$, has a wide momentum distribution, with a large spread $C^{(K)}_2$. The precise relationship between $z_x$ ($z_t$) and the second cumulants of the position (momentum) distributions will be derived in the next section. We also note that for integer filling the uncertainty relationship is trivial and does not constrain the position and momentum distributions to complement each other. Instead, one can have cases where the momentum distribution and position distribution are \emph{both} sharply peaked, e.g., a filled tight-binding band that has a localized Wannier function description and is simultaneously an eigenstate of momentum (albeit with trivial momentum eigenvalue). Interestingly, this complication does not appear for the ordinary (non-modular) uncertainty relationship between position and momentum.

We have now established that both the magnitude $z_t=|\langle \Psi|T|\Psi\rangle|$ and the phase  $\arg\langle\Psi|{T}|\Psi\rangle$ of the expectation value of the translation operator can be used to study the entanglement or localization properties of a state $|\Psi\rangle$, depending on whether it is translation invariant or not. For translation invariant states, the spread of the momentum distribution vanishes for either localized or delocalized states, so $|\langle \hat{T}\rangle |$ is insensitive to localization, while  $\arg \langle \hat{T}\rangle $ can be used to make that distinction via the result in Ref. \cite{nonzero_momentum}, perhaps after twisting the boundary conditions. In contrast, states that break translation symmetry can have their entanglement and localization properties identified from the magnitude $z_t$, as we show in the following section, and as was motivated via the uncertainty relation Eq. \ref{eq: uncertainty relation 2 zx zt}.

We will explicitly illustrate how $z_t$ encodes localization properties of different types of periodic, non-translation invariant states throughout this article. We also provide evidence that the phase $\arg \langle\hat{T}\rangle$ is insensitive to flux insertion for states that break translation symmetry only when their momentum distributions are completely flat --- that is, when the states are completely localized in position space and the magnitude $|\langle \hat{T}\rangle|$ vanishes. This provides a complementary perspective on why localized states are not sensitive to twisted boundary conditions. 

A summary of the relationship between translation invariance, entanglement and localization, and the momentum distribution is schematically depicted in Fig. \ref{fig: distribution schematic}. The top row illustrates that the momentum distribution of translation-symmetric states is always a delta function, whose peak is translated by flux insertion only for LRE/delocalized states. States that have broken translation symmetry have momentum distributions which are not delta functions, and are extended in momentum space, as shown in the bottom row of Fig. \ref{fig: distribution schematic}. Of those states, the LRE/delocalized ones are still sensitive to flux insertion, since the momentum distribution can be seen to shift as flux is inserted, as in Fig. \ref{fig: distribution schematic}(d). In contrast, states which are SRE/localized and not translation-symmetric have flat momentum distributions, as shown in Fig. \ref{fig: distribution schematic}(c), so any shift due to flux insertion goes undetected. We discuss the effects of flux insertion in Section \ref{sec: flux insertion}.

We devote the remainder of the article to states that break translation symmetry, and precisely show under which conditions $z_t$ and $z_x$ may be used to probe the localization properties of those states. In the next section we use the characteristic functions formalism to calculate the second cumulants of distributions. We apply these results to relate $z_t$ to the localization properties of states that break translation symmetry. Afterward we will move on to discuss a number of illustrative examples in one dimensional lattice models.

\section{Probability Distributions and Characteristic Functions} \label{sec: prob distr and char funcs}

As mentioned in the Introduction, states in one-dimensional periodic systems are short-range entangled when they are localized in position space, and long-range entangled when delocalized. Therefore, we can probe the entanglement of these states by determining their localization/delocalization properties. The localization of a state can be quantified via the second cumulant, or the spread, of its many-body position distribution $P(X)$. For a many-body wavefunction $\Psi$, parameterized by each particle's coordinates, the total position distribution $P(X)$ is defined as \cite{Souza_2000}
\begin{equation}
    P(X) = \langle\Psi|\delta(\hat{X}-X)|\Psi\rangle, \label{eq: position distribution}
\end{equation}
where $\hat{X}=\sum_{i}x_i\hat{n}_i$ is the many-body position operator, and $\hat{n}_i$ is the number density operator at site $i$. A key measure for localization, the \textit{localization length} $\lambda,$ is defined in terms of the second cumulant,
\begin{equation}
    \lambda^2 = \langle \hat{X}^2\rangle - \langle \hat{X}\rangle^2  \equiv C^{(X)}_2\label{eq: localization length}.
\end{equation}

A state that generates a distribution $P(X)$ that is highly peaked, such as a delta function, has a small localization length $\lambda$ compared to the system size $L$, and is thus localized. A uniform position distribution having $\lambda\gg L$, on the other hand, is highly delocalized. Furthermore, the Heisenberg uncertainty principle states that the spread in the momentum distribution $P(K)$ has a lower bound inversely proportional to the spread of the position distribution. Therefore, a highly localized state in position space has a momentum distribution with a large spread, i.e., large second cumulant $C^{(K)}_2$. The caveats when the boundary conditions are periodic and $P(X)$ is a periodic distribution will be addressed in Section \ref{sec: resta formula}.

Thus, we can study the localization, and consequently entanglement, properties of a one-dimensional many-body ground state by quantifying the second cumulants of its position and, as we will see, momentum distributions. To this end, we employ the \textit{characteristic functions} formalism, which provides a systematic way to obtain the cumulants of a distribution from the expectation values of a set of appropriate operators. The rest of this section is dedicated to reviewing this formalism, and applying it to both the position and the momentum distributions.

\subsection{Characteristic functions and localization}

We now review the characteristic functions formalism to study probability distributions. The advantage of this approach is that all the moments and cumulants of probability distributions ---and consequently the distributions themselves--- can be obtained from the characteristic functions \cite{Souza_2000, hetenyi-generating_reference}, which take the form of expectation values of certain operators for the probability distributions describing many-body states.

The characteristic functions of a generic probability distribution $P(r)$, where $r$ is a continuous real random variable, are given by the Fourier transform
\begin{equation}
    f(s)=\int_{-\infty}^{\infty}dre^{isr}P(r).
\end{equation}
The moments $M^{(r)}_n$ and cumulants $C_n^{(r)}$ of the probability distribution $P(r)$ can be obtained from the characteristic functions as
\begin{align}
    M^{(r)}_n &= \frac{1}{i^n}\frac{\partial^n f(s)}{\partial s^n}\bigg|_{s=0}, \label{eq: moment continuous}\\
    C^{(r)}_n&=\frac{1}{i^n}\frac{\partial^n \ln f(s)}{\partial s^n}\bigg|_{s=0}. \label{eq: cumulant continuous}
\end{align}
The last equation motivates calling $\ln f(s)$ the \textit{generating function} of the cumulants. The set of all the moments or all the cumulants completely determines the distribution function \cite{Souza_2000,hetenyi-generating_reference}. If the characteristic functions for all $s$ are known, the probability distribution can also be recovered from the inverse Fourier transform. 

The characteristic functions of the many-body position distribution $P(X)$, as defined in Eq. \ref{eq: position distribution}, are given by the expectation values
\begin{equation}
    f(k)=\int_{-\infty}^{\infty}dXe^{ikX}P(X) = \langle e^{ik\hat{X}}\rangle, \label{eq: characteristic position}
\end{equation}
where $k$ has the interpretation of center of mass momentum. Then the localization length $\lambda$ can be obtained from these functions through the second cumulant, using the definitions in Eqs. \ref{eq: localization length} and \ref{eq: cumulant continuous} to write
\begin{equation}
    C_2^{(X)} = - \frac{\partial^2 \ln f(k)}{\partial k^2}\bigg|_{k=0} = \lambda_x^2. \label{eq: second cumulant continuous}
\end{equation}

\subsection{Review: proof of Resta's formula} \label{sec: resta formula}

As mentioned in the introduction, we are interested in studying \textit{periodic} one-dimensional systems, where localization properties are a good proxy for entanglement properties. A system with length $L$ and $N_e$ electrons having periodic boundary conditions has eigenstates $\Psi$ that obey
\begin{equation}
\Psi(x_1,...,x_i,...,x_{N_e})=\Psi(x_1,...,x_i+L,...,x_{N_e})
\end{equation}
for any variable $x_i$ separately \cite{Resta-98-operator}, where $x_i$ is the position coordinate of the $i^\text{th}$ particle.

In periodic systems, the state $\hat{X}|\Psi\rangle$ is ill-defined, since it is not in the Hilbert space of periodic states, which makes Eq. \ref{eq: localization length} for the localization length also ill-defined \cite{Souza_2000}. Nevertheless, the state $e^{ik\hat{X}}|\Psi\rangle$ can still be well-defined in periodic systems, and hence the characteristic function $f(k)$ can also be well-defined. Therefore, the moments and cumulants of the distribution $P(X)$ are still meaningful for the localization properties of the state. This conclusion was an important contribution of Resta and Sorella in Ref. \cite{Resta-99}, and was later expanded upon using the generating function formalism by Souza, Wilkens, and Martin (SWM) in Ref.\cite{Souza_2000}. In particular, the localization length is still meaningfully identified with the second cumulant of the many-body position distribution, $\lambda_{x}^{2} = C^{(X)}_2$. 

An important aspect of this result by Resta, Sorella and SWM is that, in the thermodynamic limit, the second cumulant $\lambda_{x}^2$ can be written as a function of only the characteristic function $f(k=\frac{2\pi}{L})$ \cite{Resta-99,Souza_2000}. Recognizing that $f(\frac{2\pi}{L})=\langle e^{i\frac{2\pi}{L}\hat{X}}\rangle$, and defining $z_x\equiv |\langle e^{i\frac{2\pi}{L}\hat{X}}\rangle|$, their result can be stated as 
\begin{equation}
  z_x= e^{-\frac{1}{2}\left(\frac{2\pi}{L}\right)^2\lambda_{x}^2},\label{eq: Resta's formula}
\end{equation}
which we called \emph{Resta's formula} in the Introduction. We now review the derivation of this equation and its regime of validity, as we will follow similar steps to prove an analogous relationship for the many-body momentum distribution. Moreover, the analogous result for the momentum distribution, which relates the expectation value of the translation operator to the localization properties of a many-body state, is one of our main formal results.

In order to recover Resta's formula, it is important to recognize that states having position distributions $P_L(X)$ that are periodic in $L$, and defined as
\begin{align}
    &P_L(X)= \sum_{\omega=-\infty}^\infty P(X+\omega L), &\omega\in\mathbb{Z}, \label{eq: periodic PX}
\end{align}
have allowed momentum components that can take only the values $k=\frac{2\pi}{L}q$, for $q\in \mathbb{Z}$. 
Consequently, the characteristic functions of $P_{L}(X)$ are also functions of the discrete variables $q$, as long as $L<\infty$,
\begin{equation}
    f_q = \int_0^LdX e^{i\frac{2\pi}{L}qX}P_L(X), \ \ \ \ q\in\mathbb{Z},
\end{equation}
and it can be shown that $f_q=f(\frac{2\pi}{L}q)$, where the RHS is as in Eq. \ref{eq: characteristic position} \cite{hetenyi-geometric-cumulants}.

Since $k$ now takes only discrete values, the derivatives in Eqs. \ref{eq: moment continuous} and \ref{eq: cumulant continuous} must be approximated with finite differences \cite{Souza_2000,hetenyi-generating_reference}. In particular, for the second cumulant in Eq. \ref{eq: second cumulant continuous}, one must use finite differences to approximate the continuous second derivative up to  corrections that vanish in a particular limit. This is done as follows: let $h$ be the spacing between the possible values of a variable $y$. One form of the second finite difference of a function $g(y)$ is given by 
\begin{equation}
    \frac{\partial^2g(y)}{\partial y^2}\bigg|_{y=0} = \frac{g_1 + g_{-1} - 2g_0}{h^2} + \mathcal{O}(h^2),\label{eq: second cumulant approx h2}
\end{equation}
where $g_{q}=g(y=qh)$ for $q\in\mathbb{Z}$. Then the second cumulants of the many-body position distribution correspond to setting $g(k)=\log f(k)$, with $h=\frac{2\pi}{L}$. Rewriting $k=hq=\frac{2\pi}{L}q$, we define
\begin{equation}
    Z_q\equiv f(hq) = \langle e^{i\frac{2\pi}{L}q \hat{X}}\rangle.
\end{equation}
We can finally write the second cumulant of the position distribution as
\begin{align}
    C_2^{(X)} &= \left(\frac{L}{2\pi i}\right)^2(\ln Z_1 + \ln Z_{-1} - 2\ln Z_0) + \mathcal{O}(L^{-2}),\nonumber\\
    C_2^{(X)} &= - 2\left(\frac{L}{2\pi}\right)^2 \ln |Z_1|+ \mathcal{O}(L^{-2}),\label{eq: c2 position}
\end{align}
where to get the second line we used the facts that $Z_0 = 1$ and $Z_{-1}=Z_{1}^*$.

In the thermodynamic limit, $\mathcal{O}(L^{-2})\to 0$ and we recover Resta's formula by identifying $C_2^{(X)}=\lambda_x^2$:
\begin{equation}
  |Z_1|\equiv z_x= e^{-\frac{1}{2}\left(\frac{2\pi}{L}\right)^2\lambda_{x}^2}. \label{eq: Resta z}
\end{equation}
In the examples in Section \ref{sec: numeric examples}, we will use the quantity $z_x=|\langle e^{i\frac{2\pi}{L}\hat{X}}\rangle|$ to probe localization using Resta's formula, and compare with the results obtained using $z_t$. 

A state is considered \textit{completely localized} if $z_x \to 1$ in the thermodynamic limit, and \textit{partially} localized if $0<z_x<1$ in the same limit. The notion of complete delocalization faces the subtlety that $z_x\to0$ does not clearly distinguish between $L\ll \lambda <\infty$ and $\lambda\to\infty$ for finite system sizes. We will argue that $z_t=|\langle \hat{T}\rangle|\to 1$ is a more natural definition for complete \emph{delocalization} (as compared to complete \emph{localization}) of states that {break} translation symmetry in the following subsection.

\subsection{Momentum distribution and localization}
We have so far focused on the characteristic function and second cumulant of the many-body position distribution, and recovered Resta's formula for the localization length in terms of $z_x$. Now we turn to quantifying the second cumulant of the many-body \emph{momentum} distribution, since one of our goals is to demonstrate that the localization properties of states that break translation symmetry can be captured by this quantity. 

In analogy with the procedure to find the characteristic functions of a many-body position distribution periodic in $L$, we consider the many-body momentum distribution $P(K)$, defined as
\begin{equation}
    P(K) = \langle \Psi|\delta(\hat{K}-K)|\Psi\rangle,
\end{equation}
where the total momentum is similarly defined as $\hat{K} = \sum_j k_i\hat{n}_j$. Here, $\hat{n}_j$ is the number density in momentum space, meaning how many particles have momentum $k_j=\frac{2\pi}{L}j$. Note that we do not impose any restrictions on the system size $L$. It can either be finite, in which case the many-body momentum variable $K$ takes discrete values, or we can take the thermodynamic limit, $L\to \infty$ at fixed $a$, making $K$ a continuous variable. This is in complete correspondence with the position distribution considered in the previous section, where no conditions were imposed on the lattice spacing $a$.

The next step towards obtaining the cumulants is to calculate the generating functions of the many-body momentum distribution,
\begin{equation}
    f(x) = \int_{-\infty}^{\infty} dK \ e^{iKx}P{(K)}=\langle e^{ix\hat{K}}\rangle,\label{eq: characteristic function momentum}
\end{equation}
which is now a function of position $x$. We uniformly discretize position space, so that the position variable $x$ takes only the values $x=an$, for $n\in \mathbb{Z}$. That is, the system considered is defined on a lattice with spacing $a$.

The discrete nature of the position variable $x=an$ implies that the momentum distribution is periodic in $2\pi/a$. Indeed, the discrete Fourier Transform of a many-body state  $\Psi(x=an)$, defined in the discretized position basis, gives \begin{align*}
    \tilde\Psi(K)&=\sum_{n=0}^{N-1}e^{iK(an)}\Psi(an)\\
    &=\sum_{n=0}^{N-1}e^{i(K+\frac{2\pi}{a})(an)}\Psi(an)\\
    &=\tilde\Psi\left(K+\frac{2\pi}{a}\right),
\end{align*}
for a system size $L=aN$. Consequently, the momentum distribution is also periodic, and can be expressed as \begin{equation}
    P_{a}(K)= \sum_{\omega=-\infty}^\infty P\left(K+\omega \frac{2\pi}{a}\right),
   \end{equation} 
   in analogy with Eq. \ref{eq: periodic PX} for the periodic position distribution. Then we can write the generating functions as
   \begin{equation}
       f_n = \int^{\frac{2\pi}{a}}_0dKe^{iKan}P_{a}(K) = \langle e^{ian\hat{K}}\rangle,
   \end{equation}
   where $f_n=f(an)$ as in Eq. \ref{eq: characteristic function momentum}.

Note that the operator $ e^{ix\hat{K}} = e^{ian\hat{K}}=\hat{T}_{n}$  is the translation operator for translations by $n$ lattice constants $a$. Hence, the characteristic functions of the momentum distribution are the expectation values of translations,
\begin{equation}
    \tilde Z_n = \langle T_n\rangle.
\end{equation}

The second cumulant of the many-body momentum distribution can now be written using the finite differences formula as in Eq. \ref{eq: second cumulant approx h2}, and recognizing the lattice constant $a$ as the spacing $h$ between the allowed values of position $x$ in the lattice:
\begin{align}
     C^{(K)}_2&= -\frac{2}{a^2}\log z_t + \mathcal{O}(a^2),\label{eq: c2 momentum}
\end{align}
where we define $z_t \equiv |\tilde Z_1|=|\langle T_1\rangle|$. In the continuum limit, when $a\to 0$, the above equation reduces to 
\begin{equation}
    C^{(K)}_2 = -\frac{2}{a^2}\log z_t.\label{eq: c2 momentum simple}
\end{equation}

This formula directly relates the expectation value of the translation operator to the spread of the momentum distribution. As noted in the first section, translation symmetric states are eigenstates of momentum, and their distributions are delta functions. Hence, such states always have $z_t=1$, regardless of their localization or entanglement properties. States where translation symmetry is broken, on the other hand, may have their localization properties discerned via the magnitude $z_t$. 

When translation symmetry is broken, and at incommensurate filling where $aN_e/L\notin \mathbb{Z}$, for $N_e$ the total charge of the state, we expect localized states to have flattened/uniform momentum distributions, in the sense that $C^{(K)}_2\gg \frac{2\pi}{a}$ and $z_t\to0$ in the continuum limit (see Appendix \ref{appx: uncertainty relation}). The uncertainty relation Eq. \ref{eq: uncertainty relation 2 zx zt} then implies that $z_x>0$, which means that these states are at least partially localized. In contrast, states having $z_t\to1$ in the continuum limit have highly localized momentum distributions in momentum space, $C_2^{(K)}\ll2\pi/a$, while their position distributions are wider, $C_2^{(X)}\gg a$, implying  $z_x\to0$. We say that states having $z_t\to 1$ and $z_x\to0$ in the continuum limit are \textit{completely delocalized}. Thus we have mapped the expectation value of the translation operator to the localization properties of the state even when translation symmetry is broken. This is our main result, and we study its applicability in several different lattice models in the next sections.

\subsection{Interpretation of limits}
An important difference between Eq. \ref{eq: c2 momentum simple} and its position space version is that the former is only exact in the \textit{continuum} limit $a\to 0$, while Resta's formula Eq. \ref{eq: Resta z} is exact in the thermodynamic limit $N\to\infty$. That is to be expected for the momentum space version of Resta's result, because the continuum limit increases the size $(2\pi/a)$ of momentum space without changing the density of the allowed momenta, just as the thermodynamic limit increases the size of the system by increasing the number of sites $N$ while keeping the lattice spacing $a = L/N$ constant. In other words, the continuum limit acts in momentum space just as the thermodynamic limit acts in position space.

The difference of which limit makes the formula exact is important for applications to lattice models. The thermodynamic limit is often  most naturally taken in lattice models, and $z_x$ should converge to $1$ for localized states as the system size increases. This makes Resta's formula particularly convenient for calculating the localization length in many systems with localization transitions, such as disordered and quasiperiodic systems\cite{philip_RDM_1990,aubry_andre_1980,harper_1955}.  The continuum limit, on the other hand, is often less accessible or not well-defined in those models. The Random Dimer Model (RDM)\cite{philip_RDM_1990}, for example, has random and correlated onsite potentials in the lattice that have no clear continuum generalization with the same properties. Additionally, the Aubry-Andre model \cite{aubry_andre_1980, harper_1955} is only a deep-tight binding limit of the continuum bichromatic model, which generally has other localization features in the continuum \cite{boers_bichromatic_2007,sarma_bichromatic_2017}. 

Nevertheless, we show in the following sections that $z_t$ can elucidate the localization properties of interesting lattice models with (and without) well-defined continuum limits. Indeed we find $z_t$ is even sensitive to localization transitions in models without well-defined continuum limits such as the RDM. This makes $z_t$ an enriching complementary measure for the study of localization properties of generic condensed matter systems. Additionally, as previously discussed, the localization properties of 1D periodic systems also reflect their entanglement properties. Hence, studying the role of $z_t$ in the localization of states that break translation symmetry broadens our understanding of how the expectation value of the translation operator may encode the entanglement properties of generic periodic states. Seeing the generality of the results in Ref. \onlinecite{nonzero_momentum} for translation-invariant systems in various spatial dimensions, we are optimistic that the expectation value of the translation operator will also yield results when generalized to higher dimensions, and can perhaps even reveal LRE in systems that have topological order. We will leave such studies to future work.

\section{Lattice Model Examples} \label{sec: numeric examples}

We now calculate $z_x$ and $z_t$ for the many-body ground states for different models to illustrate (i) the use of $z_t$ as a (de)localization measure, (ii) the role of the limits, and (iii) the limitations of this approach when the continuum limit is not well-defined. 

We restrict our examples to free-fermion models, where the expectation values of the modular position or translation operators in the many-body ground state can be obtained from the occupied single-particle states. Since the many-body ground state $|\Psi\rangle$ is a Slater determinant, the expectation value of an operator $\mathcal{O}$ can be obtained from the determinant of the overlap matrix $S$ as
\begin{equation}
    \langle \Psi| \mathcal{O}|\Psi\rangle = \det S, \label{eq: overlaps general}
\end{equation}
where $S_{ij} = \langle \psi_i|\mathcal{O}|\psi_j\rangle$, and $|\psi_n\rangle$ is the $n^{th}$ occupied single particle state \cite{Resta-98-operator,comp_chem_overlaps_2016}.

As a first example of the application of $z_t$ as a localization measure, we introduce a model we will call the deterministic dimer model (DDM). The name arises because of the observation in Ref.~\cite{taylor-dimer-momentum-spectrum} that the random dimer model\cite{philip_RDM_1990} exhibits its remarkable delocalization properties because a particular scattering/Fourier component of the spatial potential vanishes identically for every random instance of the potential. For the DDM we will instead set certain Fourier components of the spatial potential to zero by hand to mimic this effect in a deterministic, and tunable way. A benefit of this construction is that the DDM has well-defined continuum and thermodynamic limits, making it an useful candidate for studying the effect of the limits on the localization measures.

\subsection{Deterministic Dimer Model}\label{subsec: DDM}
The DDM is a single-orbital, one-dimensional tightbinding model in the presence of an inhomogeneous spatial potential.
As mentioned, the main feature of the DDM Hamiltonian is a spatial potential couples some, but not all, different momenta. We will focus on the case when the coefficient $V(\delta)$ of the coupling between two values of momentum $k$ and $k'$ is nonzero only for some $0<|\delta|=|k-k'|<\Delta$, where $\Delta <\pi/a$ is arbitrary and sets the range of the allowed momentum coupling (see Fig. \ref{fig:limits potential} for a visual representation). The coupling coefficients are parameterized by $\delta$ as
\begin{align}
    V(\delta) &=  \begin{cases} 
      V & \delta\in(0,\Delta)\cup (\frac{2\pi}{a}-\Delta,\frac{2\pi}{a}), \\
      0 & \text{otherwise},
   \end{cases}\label{eq: Vq}
\end{align}
where $V\in\mathbb{R}$. The Hamiltonian then reads
\begin{align} \label{eq: det-dimer}
    H_{\text{DDM}} &= \sum_k2t\cos(ka)c^\dagger_kc_k +\sum_{k,\ \delta}V(\delta)(c^\dagger_kc_{k+\delta} + h.c.).
\end{align}

\begin{figure}
    \centering
    \includegraphics[width=0.5\textwidth]{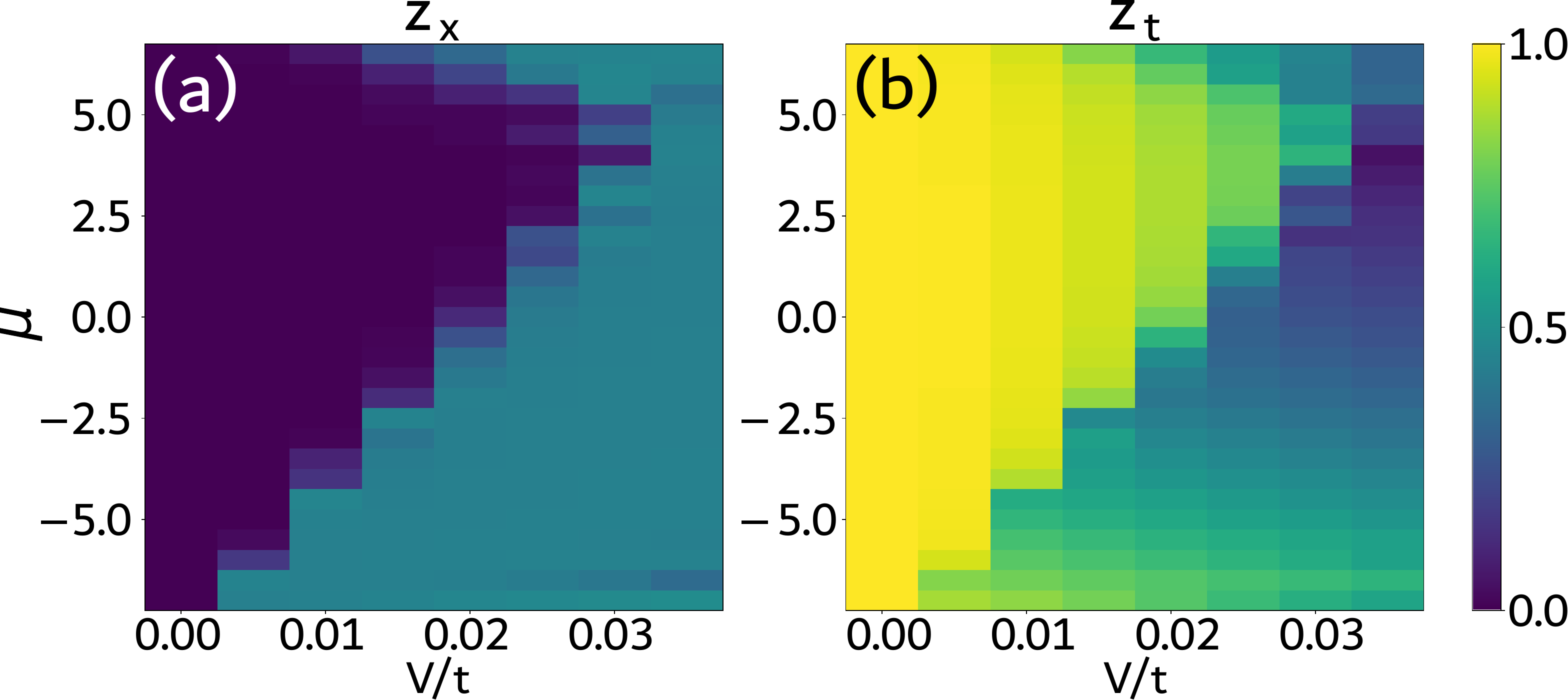}
    \caption{Localization phase diagrams for the ground state of the DDM at different chemical potentials $\mu$ and momentum coupling strengths $V/t$. Here, the total system size is $L=900a$, and $\Delta=48\frac{2\pi}{L}$. The dark-blue region in (a), for $z_x\to0$, and the bright-yellow region in (b), for $z_t\to1$, correspond to the completely delocalized phase of the DDM, and the other regions are partially localized.}
    \label{fig: DDM phase diagrams}
\end{figure}

Since for $V=0$ the system at any non-integer filling is a gapless metal, we expect the ground state of the DDM has a delocalized phase at low $V/t.$ This phase is characterized by a divergent localization length, and consequently $z_x\to 0$. As $V/t$ increases, we expect the ground state to tend toward localization. To confirm this intuition we numerically calculate the phase diagram of the ground state of the DDM using $z_x$ and $z_t$ to probe the state's localization properties for different values of the chemical potential $\mu$ and the coupling strength $V/t$, with results shown in Fig. \ref{fig: DDM phase diagrams}. For example, in Fig. \ref{fig: DDM phase diagrams}(a) we see a dark blue region for small values of $V/t$ where $z_x\sim 0$ and hence where the model is in a delocalized phase. For each $\mu$ in the $z_x$ phase diagram there appears to be a critical value of $V/t$ where the ground-state transitions from delocalized to (at least partially) localized.

At the values of the chemical potential considered in Fig. \ref{fig: DDM phase diagrams}, the filling fraction $aN_e/L$ is not an integer, so the modular position and momentum operators do not commute, $[U,T]\neq0$. Then, from the uncertainty relation Eq. \ref{eq: uncertainty relation 2 zx zt}, a completely delocalized position distribution implies completely localized momentum distribution, and vice-versa.
Although the continuum limit has not yet been taken, Eq. \ref{eq: c2 momentum} leads us to expect that the momentum distribution is highly peaked and $z_t\to 1$ in the delocalized phase (small $V/t$), which is confirmed in Fig. \ref{fig: DDM phase diagrams}(b). The uncertainty relationship hence enforces $z_x\to 0$ in this phase. Had the filling fraction been $aN_e/L=1$, the commutation relation $[U,T]=0$ would impose no constraint between the values that $z_x$ and $z_t$ may simultaneously take, and a completely localized momentum distribution would not necessarily imply a completely delocalized position distribution.

For each value of the chemical potential $\mu$, Fig. \ref{fig: DDM phase diagrams}(a),(b) also shows that there exists a value of $V/t$ after which the state is no longer completely delocalized. Instead, $z_x \approx 0.5$ while $z_t$ approaches zero as $V/t$ increases in that parameter range. Since the maximum value of $z_x$ in this regime remains constant $z_x\approx 0.5<1$ in the thermodynamic limit, this is a partially localized phase as per our definition of partially localized states in the previous section.

\begin{figure}
    \centering
    \includegraphics[width=\linewidth]{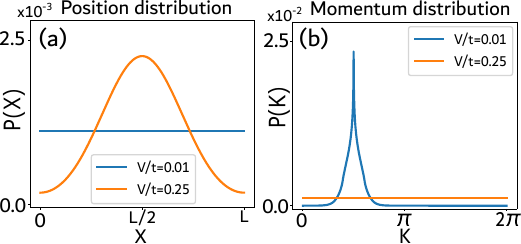}
    \caption{Position and momentum distributions for the ground state of the DDM at $\mu=0.0$. The blue $V/t=0.01$ curve is completely delocalized (flat) in position space {(a)}, while sharply peaked and completely localized in momentum space {(b)}.The orange $V/t=0.25$ curve is partially localized in position space {(a)}, and completely delocalized (flat) in momentum space {(b)}. }
    \label{fig: DDM ground state distributions}
\end{figure}

For comparison, Fig. \ref{fig: DDM ground state distributions} shows the position and momentum distributions of the ground state of the DDM at $\mu=0$ for two different values of 
$V/t$. The blue curves are for $V/t=0.01$ which is deep in the delocalized phase with $z_t\sim 1$ shown in Fig.\ref{fig: DDM phase diagrams}(b). For this value of $V/t$ we see a clear peak in the momentum distribution, while the position distribution is flat. Both are signals of a delocalized phase. For the orange curves in Fig. \ref{fig: DDM ground state distributions}, we are deep in the localized phase at $V/t=0.25$, where $z_t\sim 0$. Here, the position distribution is peaked (though perhaps not sharply peaked) and the momentum distribution is flat (c.f. the second row of Fig. \ref{fig: distribution schematic}).

Having established that the phase diagrams for $z_x$ and $z_t$ are compatible, one could then observe that Fig. \ref{fig: DDM phase diagrams} shows a gradient in the decreasing values of $z_t$ in the partially localized phase, while $z_x$ remains fixed in the same region. It is then natural to ask to what extent $z_x$ and $z_t$ may contain different information about the localization of the state. While complementary, each of the two measures is more sensitive to different aspects of localization in different limits. In the following subsections, we explore these subtleties using the DDM.

To compare the differences between  $z_x$ and $z_t$ as localization measures, it is helpful to first detail the localization properties of the single-particle eigenstates of the DDM. Then we explain how the localization of the many-body ground state may be inferred from the single-particle $z_t$ in a simple way \emph{for this model}, but not necessarily in general. Finally, we discuss the effect of the thermodynamic and continuum limits for both $z_x$ and $z_t$.

\subsubsection{Single-particle eigenstates} \label{subsec: DDM single-particle}

We now describe the single-particle eigenstates of the DDM, whose main properties can be understood from first-order perturbation theory in the strength of $V$. Note that, without the momentum coupling term, the Hamiltonian in Eq. \ref{eq: det-dimer} is diagonal in momentum space. As such, its single-particle eigenstates are plane waves with momentum $k$, denoted $|\phi_k\rangle$. For $k\notin \{0,\pi/a\}$, the plane waves with opposite momentum, $|\phi_k\rangle$ and $|\phi_{-k}\rangle$, are degenerate. 

When the momentum coupling term is turned on, the eigenstates of $H_{DDM}$ having $|k|>\Delta/2$ or $|k-\pi/a|>\Delta/2$ are not directly coupled by the potential, and do not have their degeneracy lifted at first order. Instead, the perturbed wavefunctions are modified from being single plane waves with definite momentum $k$, to being superpositions of plane waves with momenta in the vicinity of $k$. The parameter $\Delta$ determines the span of the momenta $k'$ of the plane waves in the superposition, i.e, $k'$ satisfies $k-\Delta<k'<k+\Delta$ in the perturbative regime. To first order in non-degenerate perturbation theory, the plane waves that do not have their degeneracy lifted become (not normalized)
\begin{align}
    |\psi_{(k)}\rangle=|\phi_k\rangle&+ \frac{V}{t}\sum_{\delta=\frac{2\pi}{L}}^{\Delta}\left[\frac{1}{\cos(ka)-\cos((k+\delta)a)}|\phi_{k+\delta}\rangle \right. \nonumber\\
   & +\left. \frac{1}{\cos(ka)-\cos((k-\delta)a)}|\phi_{k-\delta}\rangle\right ].\label{eq: ddm eigenstate psi}
\end{align}

\begin{figure}
    \centering
    \includegraphics[width=\linewidth]{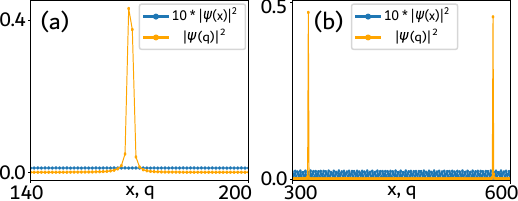}
    \caption{Position (blue) and momentum (orange) distributions for single-particle eigenstates of the DDM at $V/t=0.021$. (a) Eigenstate of the type $|\psi_{(k)}\rangle$ at energy $E = 2.99$,  whose degeneracy was not lifted by the perturbation. (b)Eigenstate of the type $\ket{\tilde\psi_{|k|}}$ at energy $E=-5.13$, whose degeneracy was lifted by the perturbation.}
    \label{fig: DDM single patcl distr}
\end{figure}

The momentum space distributions of the eigenstates of the form $|\psi_{(k)}\rangle$ are not delta functions, but peaks centered at momentum $k$ that have a finite spread. Nevertheless, for the ranges of the parameters $V/t$ considered here, the coefficients of the adjacent plane waves $|\phi_{k\pm\delta}\rangle$ are small enough that the perturbed single-particle states of the form $|\psi_{(k)}\rangle$ are narrowly peaked in momentum space, in the sense that the single-particle expectation value $z_t=|\langle\psi_{(k)}|T|\psi_{(k)}\rangle|\approx 1$. The distribution of one such state of the form of $|\psi_{(k)}\rangle$ is shown in Fig. \ref{fig: DDM single patcl distr}a, for $V/t=0.021$ and $\Delta=48\times 2\pi/L$. Note that a single peak centered at momentum $k\approx 165 \frac{2\pi}{L}$ is seen. If this distribution was a simple delta function, the magnitude of this peak would be $1$. However, the peak is smeared by the momentum coupling term, such that its maximum magnitude is less than $0.5$. 

In contrast, the unperturbed plane waves having momentum $0<|k|<\Delta/2$ or $0<|k-\pi/a|<\Delta/2$ have a slightly different fate. Plane waves having momentum in that range are directly coupled to their degenerate pairs $|\phi_{-k}\rangle$ by the perturbation, since $|k-(-k)|=|2k|<\Delta$. In this case, the degeneracy is lifted at first order in perturbation theory. The originally degenerate eigenstates take the form $\frac{1}{\sqrt{2}}(|\phi_k\rangle\pm|\phi_{-k}\rangle)$, and the perturbation further smears the two peaks in the momentum distributions of these eigenstates by coupling them to other momenta. Then the newly non-degenerate eigenstates become equal-weight superpositions of the form:
\begin{equation}
    |\tilde{\psi}_{|k|}\rangle= b_+ |\psi_{(k)}\rangle+ b_-|\psi_{(-k)}\rangle, \label{eq: superposition}
\end{equation}
where $|b_\pm|=1/\sqrt{2}$.
Then the $|\tilde{\psi}_{|k|}\rangle$ states have momentum distributions with two smeared peaks, centered at opposite momenta $\pm k$. We see such a state in Fig. \ref{fig: DDM single patcl distr}b, where the pair of momentum peaks does not have height $0.5$ exactly because of their small width. The weight of the distribution is still evenly distributed between the two peaks.

Since the unperturbed energy is given by $E=2t\cos (ka)$, first order in perturbation theory predicts an equal number $\Delta/(2\pi/L)$ of states in the lower and the higher end of the spectrum to be of the form $|\tilde{\psi}_{|k|}\rangle$. However, in the parameter range considered in this work,  $V/t\in[0.000,0.035]$, the DDM already shows some features that go beyond first order in perturbation theory. One of them is that the number of states of the form $|\tilde{\psi}_{|k|}\rangle$ is not fixed to $\Delta/(2\pi/L)$, as will be evident in the next section, when we discuss Fig. \ref{fig: DDM single particle}b. In Appendix \ref{appx: details of PT DDM}, we show that for smaller values of $V/t$, deep in the perturbative regime, we obtain exactly the expected number of $|\tilde{\psi}_{|k|}\rangle$ states on both edges of the spectrum. We also discuss further details of the departure from perturbation theory in the Appendix. In the main text, we choose the parameter range $V/t\in[0.000,0.035]$, slightly beyond a perfect description by first order perturbation theory, because it illustrates different aspects of the localization transition in the DDM more explicitly.

Now that we have some understanding of the types of single-particle states we expect to find the in the DDM we are ready to discuss the localization properties of the single-particle states as a function of $V/t$. Furthermore, we will argue that these properties may be used to simply infer the many-body localization properties of the many-body ground state of the DDM we computed in Fig. \ref{fig: DDM phase diagrams}.

\subsubsection{Single-particle and many-body state localization}\label{DDM subsec 2}

\begin{figure}
    \centering
    \includegraphics[width=1\linewidth]{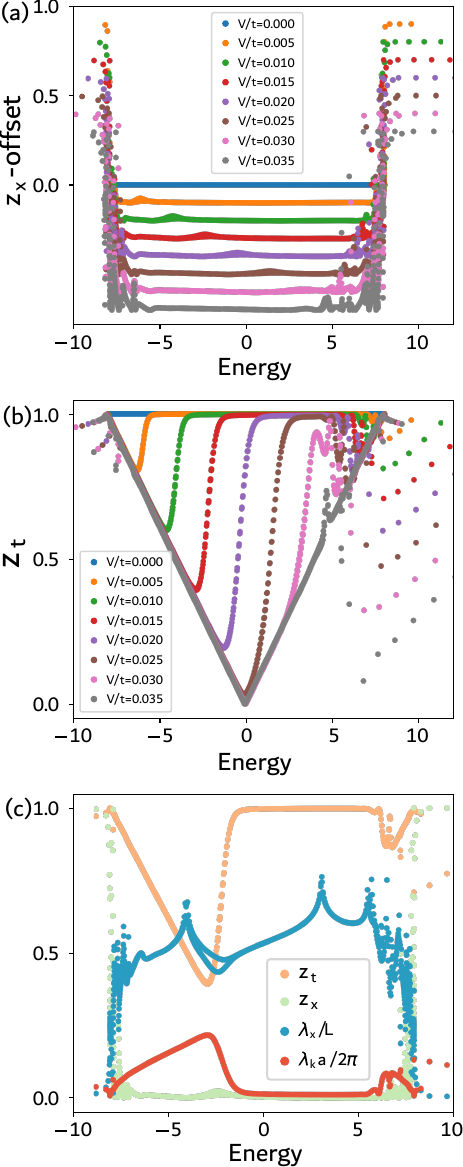}
    \caption{Localization measures for each single-particle state, arranged by energy, at different momentum hopping strengths $V/t$, for $L=900a$. (a) Single-particle $z_x$ for the eigenstates of the DDM. Each value of $V/t$ is plotted with a negative vertical offset for clarity. (b) Single-particle $z_t$ for the eigenstates of the DDM. (c) Comparison between the single-particle $z_x$, $z_t$, and the localization lengths in position, $\lambda_x$, and momentum space, $\lambda_k$, for $V/t=0.015$.}
    \label{fig: DDM single particle}
\end{figure}

For certain models, probing the localization properties of the single-particle states using $z_t$ can reveal deeper insights about the many-body ground states. This is the case for the DDM, whose localization phase diagram, we show, is more directly inferred from the single-particle localization properties determined by $z_t,$ than the single-particle localization properties given by $z_x$. To show this, we first plot in Fig. \ref{fig: DDM single particle} the values of $z_x$ (\ref{fig: DDM single particle}a) and $z_t$ (\ref{fig: DDM single particle}b) for each single-particle eigenstate of the DDM, indexed by their energies, for different values of the relative coupling strength $V/t$ at fixed $\Delta=48 \times \frac{2\pi}{L}$. 

The states of the form $|\tilde\psi_{|k|}\rangle$, which are directly coupled by the potential are expected to be present at the lower and upper ends of the spectrum. Since these states are strongly coupled by the potential we would expect them to be more strongly localized. Such states have 
\begin{align}
z_t(k)&=\langle\tilde\psi_{|k|}|T|\tilde\psi_{|k|}\rangle \nonumber\\&= \sqrt{1-4|b_+|^2(1-|b_+|^2)\sin^2(ak)}.\label{eq: zt v function}
\end{align}
For $|b_+|\approx|b_-|\sim\frac{1}{\sqrt{2}}$, this yields $z_t(k)\approx|\cos(ka)|$, which corresponds to a ``V"-shaped curve as a function of energy. In Fig. \ref{fig: DDM single particle}b, we can identify the states of the form $|\tilde\psi_{|k|}\rangle$ as those which follow that curve. Indeed, for each value of $V/t\lesssim 0.04$, there exists an energy $E_{V,1}$ below which $z_t$ follows the ``V"-shaped curve. At the very upper end of the spectrum, this also happens to a smaller degree, after some energy $E_{V,2}>E_{V,1}$ (the asymmetry between the lower and upper ends of the spectrum is discussed in Appendix \ref{appx: details of PT DDM}). For example, at $V/t=0.015$, the red curve in Fig. \ref{fig: DDM single particle}b, $E_{V,1}\approx -3$ and $E_{V,2}\approx 7$. Note that $E_{V,1}$ increases with $V/t$. Then a larger proportion of eigenstates has $z_t(k)=|\cos(ka)|$ at higher $V/t.$ Heuristically, such states, having $z_t<1$ are more localized, and this will have consequences for $z_t$ of the many-body ground state of this model, as discussed in the next subsection.

The single-particle eigenstates between energies $E_{V,1}$ and $E_{V,2}$ have $z_t\approx 1$ because their single-particle momentum distributions peaked at a single value of momentum, similar to Fig. \ref{fig: DDM single patcl distr}a. Hence, these states are of the form $|\psi_{(k)}\rangle,$ as expected, since they lie in the range of the band where the states would not be directly coupled by a perturbatively weak potential. For stronger potentials, e.g., when $V/t > 0.04$, the momentum coupling in $H_{DDM}$ (Eq. \ref{eq: det-dimer}) dominates and we cannot identify a clear $E_{V,1}$ after which the form of the eigenstates qualitatively changes from $\ket{\tilde\psi_{|k|}}$ to $\ket{\psi_{(k)}}$. Instead, all eigenstates in this regime are superpositions of plane waves with opposite momentum, having the form of $|\tilde\psi_{|k|}\rangle$ in Eq. \ref{eq: superposition}. This is reflected in Fig. \ref{fig: DDM single particle}b by the $V/t=0.035$ curve in gray, which is almost completely in the ``V"-shape.

In order to clearly visualize to what degree of localization each value of $z_x$ and $z_t$ corresponds, we also calculate the localization lengths for each eigenstate in position and momentum space, shown in Fig \ref{fig: DDM single particle}c. We take the localization lengths in position and in momentum space to be
\begin{align}
    &\lambda_x = \sqrt{C^{(X)}_2},&\lambda_k = \sqrt{C^{(K)}_2},
\end{align}
respectively. We plot in Fig. \ref{fig: DDM single particle}c the localization lengths normalized by the span of the space where the distributions are defined, $L$ for the position distribution and $2\pi/a$ for the momentum distribution. The smaller the ratio of the localization length to the size of the space of the distribution, the more concentrated, or less spread, the probability distribution is for that single-particle eigenstate.

To see how the information in $z_t$ is illuminating, choose a fixed chemical potential line in Fig. \ref{fig: DDM phase diagrams}, say $\mu=0$. Comparing with the single-particle state $z_t$ diagram in Fig. \ref{fig: DDM single particle}b, it is clear that only for $V/t< 0.02$ do we have completely delocalized single-particle states filled at that chemical potential. For higher $V/t$ the filled single-particle states are at most partially delocalized, and the many-body ground state becomes more localized, consistent with the phase diagram. Thus, from the single-particle $z_t$ information we can seemingly correctly infer the phase information in the many-body phase diagram.

While this strategy works from the single-particle $z_t$ data in Fig. \ref{fig: DDM single particle}b, it is not as clear how to carry this out from the $z_x$ data for single-particle states in Fig. \ref{fig: DDM single particle}a, or the position localization length in Fig. \ref{fig: DDM single particle}c. Although Fig. \ref{fig: DDM single particle}a shows a small local peak in $z_x$ at the same value of energy where $z_t$ increases for each $V/t$, the regions before and after those peaks are close to zero except at the very edges of the energy spectrum. With only that information, it would be difficult to infer that only after filling the higher energy single-particle states would the many-body ground state become delocalized. 

Even the spatial localization length data in Fig. \ref{fig: DDM single particle}c does not make the situation clearer. Although the local $z_x$ peak corresponds to a local minimum of the localization length, with $\lambda_x/L<0.5$, the values of $\lambda_x/L$ on either side of this local minimum are higher than $0.5$ in comparable ways. Then from the $z_x$ and $\lambda_x/L$ plots alone, it is not clear that the single-particle states with energy higher than that of the most localized ones are completely delocalized, in the sense that their momentum distribution is highly peaked and $z_t\to 1$. Therefore, it would be difficult to infer from $z_x$ and $\lambda_x/L$ that filling the delocalized single-particle eigenstates after the most localized ones would make the many-body ground state delocalized as Fig. \ref{fig: DDM phase diagrams}a shows. 

Nevertheless, both the $z_x$ and $z_t$ plots for the \emph{many-body} ground state of the DDM, presented in Fig. \ref{fig: DDM phase diagrams}, show that, at values of $V/t$ and $\mu$ where the \emph{single-particle} states having $z_t\to1$ are filled, the many-body ground states are delocalized. Interestingly, for this model, the second cumulant of the momentum distribution of the single-particle eigenstates, captured by $z_t$ and $\lambda_k$, provides clearer information about the localization properties of this model compared to the second cumulant of the position distribution, which is typically used to study localization. As we will see below, this conclusion is model dependent, and some models will likely have ground-state localization properties that are more easily inferred from single particle $z_x$ data instead of $z_t$ data.

To understand this outcome we need to understand why the single-particle localization measure $z^{sp}_t$ better reflects the many-body ground state's $z^{MB}_t$ than the similar quantities for $z_x$. Recall that the many-body $z^{MB}_t$ is the magnitude of the determinant of the matrix of overlaps of the filled single-particle states with the translation operator, so Eq. \ref{eq: overlaps general} gives
\begin{align}
    z^{MB}_t=|\langle \Psi|\hat{T}|\Psi\rangle|=|\det S|,
\end{align}
for $S_{ij}=\langle\psi_i|\hat{T}|\psi_j\rangle$ and $\{|\psi_n\rangle\}$ the set of the occupied single-particle eigenstates in the ground state. The diagonal entries of $S$ are the single-particle $z_t^{sp}$ for each occupied single-particle eigenstate, and one of the terms of the determinant $\det S$ is the product of the diagonal entries, $\prod_{n\in \text{occupied}}z^{sp,(n)}_t$. The remaining terms have factors which are overlaps between different single-particle states, $\langle\psi_i|T|\psi_{j\neq i}\rangle$.

We claim that the overlaps of the form $\langle\psi_i|T|\psi_{j\neq i}\rangle$, while generally nonzero, are much smaller than the diagonal entries for the eigenstates of the DDM, and the determinant $\det S$ is then dominated by $\prod_{n\in \text{occupied}}z^{sp,(n)}_t$. The reason is that, for the values of $V/t$ that we study, the eigenstates of the DDM are superpositions of a finite number of plane waves with adjacent momenta. Then the overlaps $\langle\psi_i|T|\psi_{j\neq i}\rangle$ are  non-negligible compared to $\langle\psi_i|T|\psi_{ i}\rangle$ for only a small set of states $|\psi_j\rangle$ having immediately adjacent momentum peaks to $|\psi_i\rangle$. Therefore, the product of $z_t$ for the single-particle eigenstates dominates the many-body $z_t^{MB}$ in the DDM. This justifies using the values of the filled single-particle $z_t^{sp}$, as shown in Fig.\ref{fig: DDM single particle}b, to estimate the many-body $z_t^{MB}$ at each chemical potential to compare with phase diagram in Fig. \ref{fig: DDM phase diagrams}b. 

At first glance, it might seem that the partially-localized single-particle states, having $z_t^{sp}<1$, would inevitably lower the value of $z_t^{MB}$ away from $1$, even when the chemical potential is set at an energy of completely delocalized states, e.g., $E_{V,1}\leq\mu<E_{V,2}$. We argue that this is not the case for this model. The argument is based on the observation that a generic unitary transformation acting on only the occupied single-particle orbitals will not change the magnitude of the determinant of the overlap matrix $S.$ In this model, such a unitary transformation can be used to effectively decouple the partially-localized occupied energy eigenstates into a basis of pure plane waves, each having $z_t^{sp}=1.$ This is a momentum-space version of the concept that a many-body state for a completely full, single-band tight-binding model is actually localized even though it is built from delocalized plane waves, i.e., a unitary transformation on the complete set of plane waves can decouple them into local, completely-onsite orbitals.  

For example, consider the partially localized states $|\tilde\phi_{\pm}(k)\rangle=\frac{1}{\sqrt{2}}(|\phi_k\rangle\pm |\phi_{-k}\rangle)$. These states are of a similar form to the perturbative eigenstates in the DDM that are directly coupled by the potential. Moreover, they can be obtained from a unitary transformation $W$ applied to the set of plane waves $\{|\phi_{\pm k}\rangle\}$, which span a subspace of the eigenspace of the translation operator $\hat T$. Hence, the set $\{|\tilde\phi_{\pm}(k)\rangle\}$ spans the same subspace. Individually, each state $|\tilde\phi_{\pm}(k)\rangle$ has $z_t^{sp}=|\cos(ka)|$, which is generically less than unity. If both states are filled in the ground state we will demonstrate that $z_t^{MB}=1$, i.e., it is not equal to $\cos^2{(ka)}$ as one might naively guess. 

To show this, consider the overlap matrix $S$ of the two occupied states $\{|\tilde\phi_{\pm}\rangle\}$,
\begin{align}
    S &= \begin{pmatrix}
        \langle\tilde\phi_{+}|\hat T|\tilde\phi_{+}\rangle &\langle\tilde\phi_{+}|\hat T|\tilde\phi_{-}\rangle\\
        \langle\tilde\phi_{-}|\hat T|\tilde\phi_{+}\rangle & \langle\tilde\phi_{-}|\hat T|\tilde\phi_{-}\rangle
    \end{pmatrix},\\
    &= \begin{pmatrix}
        \langle \tilde\phi_+|\\\langle\tilde \phi_-| 
    \end{pmatrix} \hat{T}\begin{pmatrix}
        | \tilde\phi_+\rangle&
        | \tilde\phi_-\rangle
    \end{pmatrix}.
\end{align}
Notice that $S$ is the matrix representation of the translation operator $\hat{T}$, written in the $\{|\tilde\phi_{\pm}\rangle\}$ basis, and restricted to the subspace spanned by those states. We can perform a basis transformation from $\{|\tilde\phi_\pm\rangle\}$ to the plane wave eigenbasis $\{|\phi_{\pm k}\rangle\}$ via the inverse unitary transformation, obtaining another matrix
\begin{align}
    S_W&= \begin{pmatrix}
        \langle \tilde\phi_+|\\\langle\tilde \phi_-| 
    \end{pmatrix} W^{-1}\hat{T}W\begin{pmatrix}
        | \tilde\phi_+\rangle&
        | \tilde\phi_-\rangle
        \end{pmatrix}\label{eq: Sw}\\
        &=\begin{pmatrix}
        \langle \phi_k|\\\langle \phi_{-k}| 
    \end{pmatrix} \hat{T}\begin{pmatrix}
        | \phi_k\rangle&
        | \phi_{-k}\rangle
        \end{pmatrix}\\
        &= \begin{pmatrix}
            e^{ik}&\\
            &e^{-ik}
        \end{pmatrix}.
\end{align} 
The matrix $S_W$ is the translation operator written in the eigenbasis $\{|\phi_{\pm k}\rangle\}$, and can be obtained from $S$ by the diagonalization $S_W=WSW^{-1}$. Since $W$ is a unitary transformation, $|\det S| = |\det S_W|= 1$. Remarkably, from the form of our model, this decoupling idea applies for most of the partially localized occupied states, and hence these sets of states do not diminish $\vert \det S\vert$. Using this idea, we argue in Appendix \ref{appdx: ddm z=1} that setting $\mu= E_{V,1}$ in the DDM produces $z_t^{MB}\approx 1$, even though many filled single-particle states are partially localized.

Conversely, should the chemical potential be set at $\mu<E_{V,1}$, only a subset of those partially-localized states will be filled. Then $|\det S|<1$, as $\det S$ will be dominated by the product of the $z_t^{sp}<1$ of the filled partially-localized states, i.e.,  those that cannot be decoupled into plane waves by acting only in the space of occupied states. The details of this argument are shown in Appendix \ref{appdx: ddm z=1}. Therefore, we can use the product $\prod_{n\in \text{occupied}}z^{sp,(n)}_t$ as a proxy for $z_{t}^{MB}$ for the ground state of the DDM, and both the partially-localized and the localized phases are accurately captured by this estimate.

In contrast, for the DDM we do not expect there to be a clear connection between the single-particle $z_x^{sp}$ and the many-body ground state quantity $z_x^{\text{MB}}$. The single-particle eigenstates of the DDM are at most partially localized and extend over large regions of position space, with the exception of the states at the edges of the spectrum, as illustrated by $\lambda_x/L$ in Fig.\ref{fig: DDM single particle}c. Hence, the matrix with elements $S_{ij}=\langle\psi_i|e^{i\frac{2\pi}{L}\hat{X}}|\psi_j\rangle$ has off-diagonal elements that are not generally dominated by the diagonal entries. Thus, $\det S$ cannot be well-approximated by just the term with the product of the diagonals of $S$, i.e., it is not well-approximated by considering only the single-particle $z_x$ values. 

We conclude that, in models such as the DDM where the single-particle eigenstates are mostly localized in momentum space, the single-particle $z_t$ offers a better probe for the localization properties of the many-body ground state than the single-particle $z_x$. Conversely, if a model has single-particle eigenstates that are tightly localized in position space, the many-body $z^{MB}_x$ would reflect the single-particle $z^{sp}_x$ more than the single-particle $z^{sp}_t$. We remark that although they did not play much of a role in our discussions thus far, the localization properties of the single-particle states at the lowest and highest energies of the spectrum, which have \emph{both} $z_x\approx 1$ and $z_t\approx1$ simultaneously, have interesting properties allowed by the parametrically weak commutation relation, and we look at them in slightly more detail in Appendix \ref{appx: single-particle localization}.

\subsubsection{Continuum and thermodynamic limits}

Having studied the localization properties of the single-particle eigenstates and the many-body ground state of the DDM, we now investigate the role of the continuum and thermodynamic limits on the accuracy of $z_x$ and $z_t$ as (de)localization measures. The DDM has well-defined continuum and thermodynamic limits we describe below, making it a good model to study the effects of either limit on the localization measures $z_x$ and $z_t$.

\begin{figure}
    \centering
    \includegraphics[width=0.8\linewidth]{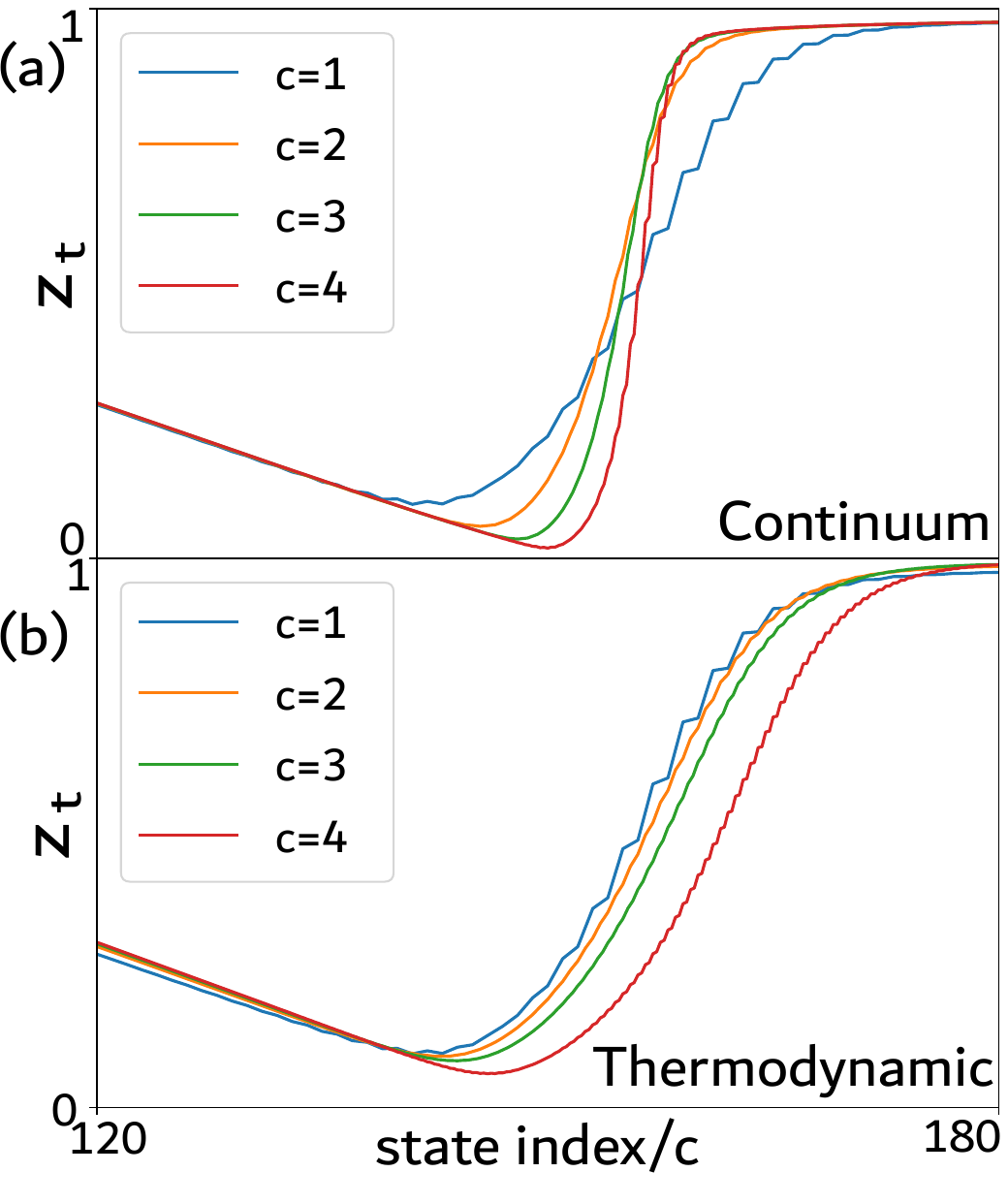}
    \caption{Single-particle $z_t^{sp}$ of a range of eigenstates of the DDM in the continuum (a) and thermodynamic (b) limits. The plots show $z^{sp}_t$ for states between indices $120$ and $180$, in increasing order of energy, at $V_0/t_0=0.071$. The lower index states in these plots have two peaks in their momentum distributions, and the higher index states are approximate delta functions in momentum space. (a) The continuum limit makes the $z_t^{sp}$ increase more sharply to $1$ for the completely delocalized eigenstates (states on the RHS of the energy range). (b) The thermodynamic limit shifts the $z_t^{sp}$ curve, making $z_t^{sp}$ smaller for the delocalized eigenstates (especially the delocalized states near the transition).}
    \label{fig: DDM single particle limits}
\end{figure}

\begin{figure*}
    \centering
    \includegraphics[width=0.6\linewidth]{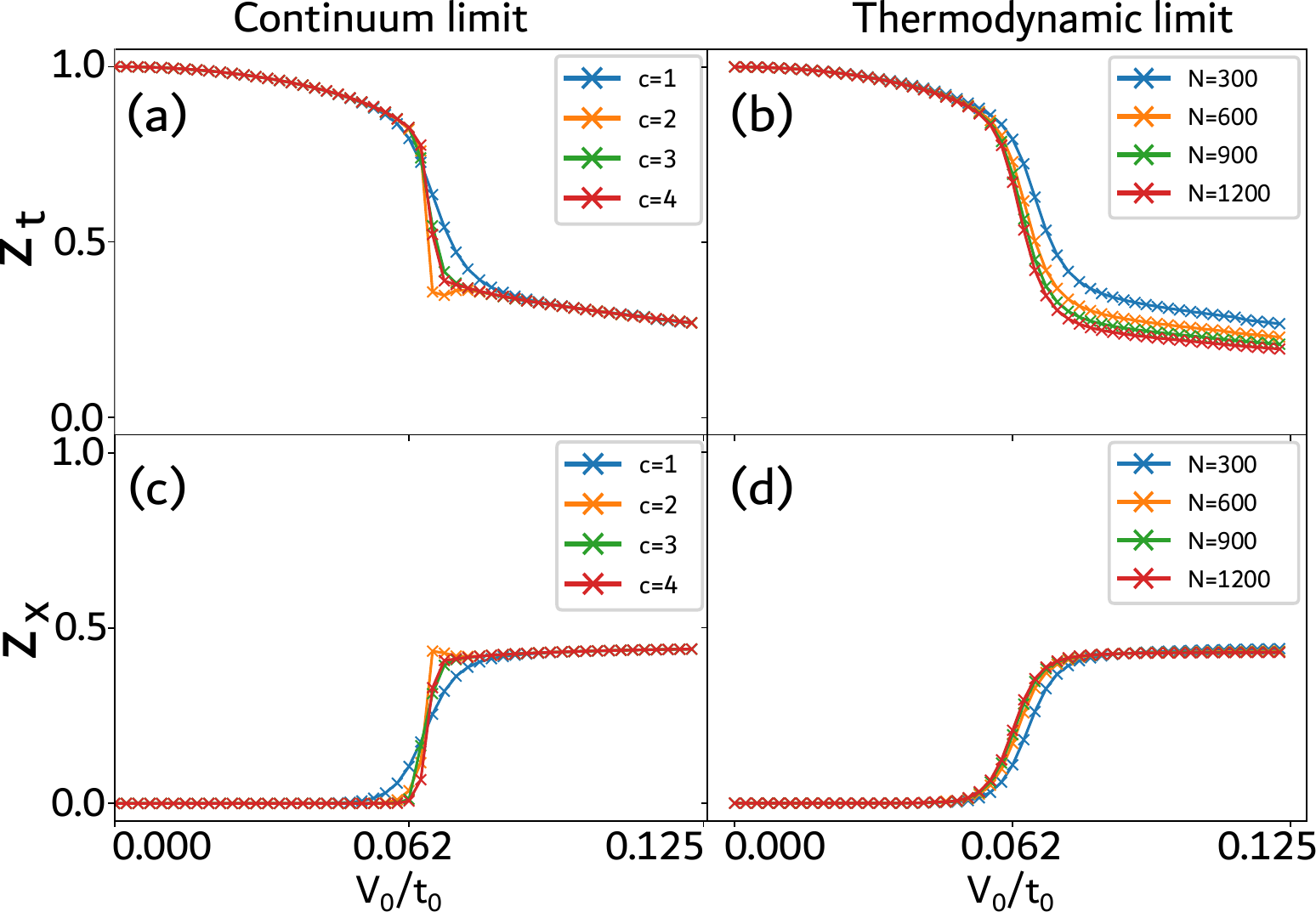}
    \caption{Effects of the continuum (a, c) and thermodynamic (b, d) limits on $z_t$ (a, b) and $z_x$ (c, d) for the many-body ground state of the DDM at $\mu=0$. (a,c) Higher scaling-factor $c$ corresponds to decreasing lattice spacing $a$, and moving towards the continuum limit. The transition from the delocalized to the partially localized phase in both $z_t^{MB}$ (a) and $z_x^{MB}$ (c) becomes sharper in the continuum limit. (b, d) The thermodynamic limit decreases the value of $z_t^{MB}$ in the partially localized phase (b), and makes the transition in $z_x^{MB}$ happen at smaller values of $V_0/t_0$ for larger system sizes.}

    \label{fig: limits}
\end{figure*}

As shown in Section \ref{sec: prob distr and char funcs}, $z_x$ and $z_t$ are exact exponentials of the second cumulant of their respective distributions only in the thermodynamic and continuum limits, respectively (see Eqs. \ref{eq: c2 position} and \ref{eq: c2 momentum}). Therefore, we expect that taking these limits is important for the accuracy of the phase diagram obtained using each measure. In this section, we numerically investigate the effect of taking each limit for the localization measures in the DDM, with the results shown in Figures \ref{fig: DDM single particle limits} and \ref{fig: limits}. Our numerical findings corroborate our analytic result in Eq. \ref{eq: c2 momentum simple} that $z_t\to 1$ for delocalized states as the continuum limit is taken, supporting the claim that $z_t$ detects localization transitions more sharply in that limit. We also present evidence that the continuum limit makes both $z_x$ and $z_t$ more sensitive to delocalized states, while the thermodynamic limit makes the measures more sensitive to localized states.

In order to systematically take the limits, we start with a system having lattice spacing $a_0=1$ and length $L_0=300a_0$. Then the system has $N_0=L_0/a_0$ lattice sites. The thermodynamic limit is taken by multiplying the original length $L_0$ by a constant $c\geq 1$, while keeping the lattice spacing $a_0$ constant. Hence, the thermodynamic limit amounts to replacing $L_0\to L=cL_0=a_0N$, where $N=cN_0$ is the new number of sites in the lattice. In the continuum limit, in contrast, the system size $L_0$ is left invariant, while $a_0\to a=a_0/c$. Since $L=aN=\frac{a_0}{c}N=L_0$, the number of lattice sites also increases by a factor of $c$, $N=cN_0$. However, unlike in the thermodynamic limit, in the continuum limit the density of lattice sites increases as the lattice spacing decreases. This microscopic change is reflected in the position-space hopping coefficient, which scales with the lattice spacing as $t\propto 1/a^2$ (see Appendix \ref{appx: limits} for details). Then the position hopping coefficient must also be scaled in the continuum limit as $t_0\to t=c^2t_0$.

\begin{figure}
    \centering
    \includegraphics[width=\linewidth]{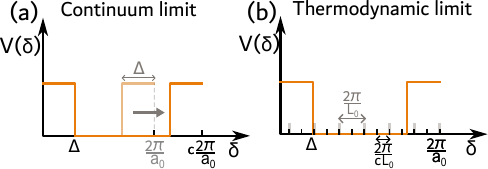}
    \caption{Effect of the continuum {(a)} and the thermodynamic {(b)} limits for the shape of the momentum coupling coefficient $V(\delta)$. {(a)} The width of the region where $V(\delta)=0$ increases as the continuum limit is taken. {(b)} In the thermodynamic limit, the values of $\delta$ sampled become more finely discretized, but the shape of $V(\delta)$ does not change.}
    \label{fig:limits potential}
\end{figure}

The procedure above for taking the thermodynamic and continuum limits is generic for 1D periodic lattice models, i.e., it will generate the appropriate limit when they are well-defined. For the specific case of the DDM, it is also important to note that $\Delta$ remains constant in either limit. Figure \ref{fig:limits potential} illustrates the effect of taking each limit on the momentum coupling function $V(\delta)$. Figure \ref{fig:limits potential}a shows that the region where $V(\delta)$ is zero becomes wider in the continuum limit, as the momentum space Brillouin zone widens as $\frac{2\pi}{a_0}\to c\frac{2\pi}{a_0}$. Meanwhile, Fig. \ref{fig:limits potential}b shows that the scale/shape of $V(\delta)$ is preserved in the thermodynamic limit, but $\delta\propto 2\pi/L$ approaches a continuous variable as the spacing between the allowed values of momentum, integer multiples of $\frac{2\pi}{L} = \frac{1}{c}\frac{2\pi}{L_0}$, decreases. Finally, the limits in the DDM must be taken in a way to ensure that the onsite potential in position space, $V(x)=\sum_{\delta=0}^{\Delta} 2 V(\delta)\cos{(\delta x)}$, has a critical value at the delocalization transition that remains constant in the thermodynamic limit. In the thermodynamic limit, $\delta$ is a continuous variable, so 
\begin{equation}
    V(x)_{\text{thermo}} = \frac{L}{2\pi}\int_0^{\Delta} 2V\cos(\delta x)d\delta. 
\end{equation}
Therefore, for any value of $V(x)$ to be independent of the system size $L$, the nonzero momentum coupling coefficient $V$ must scale as $1/L$. As an intermediate step, it is then natural to define a constant $\tilde{V}$ such that $V=\frac{a}{L}\tilde{V} = \frac{1}{N}\tilde{V}$. Since both the thermodynamic and continuum limits have $N_0\to N= cN_0$, we can further write $V=\frac{1}{N}\tilde V = \frac{1}{c}\left(\frac{1}{N_0}\tilde V\right)\equiv \frac{1}{c}V_0$. Then the original momentum hopping coefficient $V_0$ scales as $V_0\to V= V_0/c$. 

In Fig. \ref{fig: DDM single particle limits} we show the effect of taking the continuum and the thermodynamic limits for the single-particle $z_t^{sp}.$ We compute $z_{t}^{sp}$ for a range of single-particle eigenstates of the DDM around the energy $E_{V,1}$, where we previously observed the eigenstates changing from the form $\ket{\tilde\psi_{|k|}}$ to the form ${\ket{\psi_{(k)}}}$. Then, in Fig. \ref{fig: limits}, we plot the values of $z^{MB}_x$ and $z^{MB}_t$ for the many-body ground state of the DDM at half-filling for different values of $V/t$, and take either limit. In order to see the effect of taking the thermodynamic and the continuum limits, we scale the number of degrees of freedom by a factor of $c=\{1,2,3,4\}$. We plot $z_{x,t}$ as a function of $V_0/t_0$, not $V/t$, to facilitate the comparison of the localization transitions at different $c$. Recall that $\frac{V}{t}=\frac{1}{c}\frac{V_0}{t_0}$ in the thermodynamic limit, and $\frac{V}{t}=\frac{1}{c^3}\frac{V_0}{t_0}$ in the continuum limit. We scale $V$ accordingly in each limit to compare the same values of $V_0/t_0$ in the plots in Fig. \ref{fig: limits}. That is, when we use the label $V_0/t_0$ on figure axes it has two different meanings depending on if the figure shows the continuum ($V_0/t_0 = c^3 V/t$) or thermodynamic limit ($V_0/t_0=cV/t$).

With the scaling set, let us turn our focus to the contents of Figs. \ref{fig: DDM single particle limits} and \ref{fig: limits}. Recall that Eq. \ref{eq: c2 momentum} implies that a momentum distribution that is sharply localized in momentum space ($\lambda_k=\sqrt{C^{(K)}_2}\ll\frac{2\pi}{a}$) has $z_t\to1$ in the continuum limit. Then we expect the continuum limit to take $z_t$ closer to its maximum value $1$ for completely delocalized states. We already showed that the single-particle eigenstates at a given $V/t<0.04$ and energies $E_{V,1}<E<E_{V,2}$ are almost plane waves and have $z_t$ close to $1$ (see Fig. \ref{fig: DDM single particle}b). For states in this energy range we hence expect the value of  the single-particle $z_t^{sp}$ to converge to $1$ as the continuum limit is taken. In Fig. \ref{fig: DDM single particle limits}a, we show $z^{sp}_t$ for the single-particle eigenstates when $V_0/t_0=0.071$ (to compare with Fig. \ref{fig: DDM single particle} see note \footnote{To compare this value of $V_0/t_0=0.071$ with the $V/t$ values in Fig. \ref{fig: DDM single particle}c, recall that in Fig. \ref{fig: DDM single particle}c the system size is $L=900a$, which is $c=3$ times our reference system size $L_0=300a$ when taking the limits. Then Fig. \ref{fig: DDM single particle}c corresponds to taking the continuum limit with $c=3$. Using the conversion $V/t=(1/c)V_0/t_0$ for the continuum limit, we get that $V_0/t_0=0.071\to V/t=0.024$, which can be compared with the $V/t=0.025$ line in Fig. \ref{fig: DDM single particle}c.}), and take the continuum limit by decreasing the lattice spacing as $a_0\to a=a_0/c$. Figure \ref{fig: DDM single particle limits}a shows that the single-particle states for the upper energy range, which are expected to be almost delta-functions in momentum space like in Fig. \ref{fig: DDM single patcl distr}a, have an increasing $z^{sp}_t\to1$ as the limit is taken. Simultaneously, the states at lower energies, which are formed from two momentum peaks as in Fig. \ref{fig: DDM single patcl distr}b, have a lower $z^{sp}_t$ as the continuum limit is taken. Thus, the continuum limit generates a sharper contrast in the localization properties of the single-particle eigenstates computed via $z^{sp}_t$.

A sharper contrast is also observed in the localization transition of the many-body ground state at $\mu=0$, as captured by $z_t^{MB}$ in Fig. \ref{fig: limits}a. However, the contrast is not as sharp as we might have expected because $z_t^{MB}$ is not as sharply pinned to 1 for values of $V_0/t_0$ away from the transition. In fact, we argued in Section \ref{DDM subsec 2} that $z_t^{MB}\approx 1$ when $E_{V,1}\leq\mu<E_{V,2}$ because the filled single-particle states far below $\mu$ can be decoupled into plane waves, and the many-body $z_t^{MB}$ is not diminshed by the contributions of these states. We can see from Fig. {\ref{fig: DDM single particle}b} that there are many values of $V/t$ for which $E_{V,1}<\mu=0 $, and therefore we would naively expect $z_t^{MB}\approx 1$ for $V/t$ before the localization transition at $V/t\approx0.062$ in Fig. \ref{fig: limits}. However, Figs. \ref{fig: limits}a,b show that, as $V_0/t_0$ increases from zero, the maximum value of $z^{MB}_t$ decreases away from $1$ for any $c$, even before the transition. 

This is only an apparent contradiction of our previous explanation. In fact, Fig. {\ref{fig: DDM single particle}b} shows that, as $V_0/t_0$ increases in the delocalized phase, there are more partially-localized single-particle states that are filled when $\mu=0$, i.e., more partially-localized states appear in the spectrum below $E_{V,1}$. The pair of momentum peaks in the momentum distribution of these partially-localized single-particle eigenstates becomes more and more separated for states at energies closer to $E_{V,1}$. This is reflected in the fact that $z_t^{sp}$ has lower minimum values (a deeper $V$-shape) for increasing $V/t$ in Fig. {\ref{fig: DDM single particle}b}. Therefore, at higher $V_0/t_0$ and fixed $\mu=0$, not only are there more partially-localized single-particle states with smaller single-particle $z_t^{sp}$ that are filled, but also these states couple the plane waves of a wider range of momenta $k$. We argue in detail in Appendix \ref{appdx: ddm z=1}  how this wider range of plane waves means that there does not exist a unitary transformation that completely decouples the occupied, partially-localized states into plane waves Hence, we expect $z_t^{MB}$ to drop in magnitude as $V_0/t_0$ increases, even before the localization transition. In other words, the filled single-partice states can only be exactly decoupled into a set of plane waves spanning the same subspace at perturbatively low $V_0/t_0$, and as $V_0/t_0$ increases, this becomes a worse approximation, and $z_t^{MB}$ is no longer exactly equal to $1$ on the delocalized side of the transition.

As $V_0/t_0$ approaches the localization transition at $V_0/t_0\approx0.062$, the many-body $z^{MB}_t$ in Fig. \ref{fig: limits}a (for $\mu=0$) remains closer to its maximum allowed value as $c$ is increased toward the continuum limit. By ``maximum allowed value", we mean the value of $z_t^{MB}$ that would be obtained at $\mu=E_{V,1}$, where the partially-localized single-particle states cannot be linearly transformed to a set of plane waves spanning the same space. As previously discussed, this prevents $z_t^{MB}$ from being exactly $1$ when $\mu=E_{V,1}$. Let $Z$ be the maximum value of $z_t^{MB}$, obtained at $\mu={E_{V,1}}$. At larger $E_{V,1}<\mu<E_{V,2}$, the single-particle states filled after energy $E_{V,1}$ are delocalized almost plane waves. Then the total $z_t^{MB}$ is given by
\begin{equation}
z_t^{MB}=Z\times\prod^\mu_{E>E_{V,1}}z_t^{sp,(E)},
\end{equation}
where the superscript $(E)$ in $z_t^{sp,(E)}$ indicates that this is the $z_t^{sp}$ for the single-particle state having energy $E$. Since these states at energy $E_{V,1}<E<\mu$ are almost plane waves, $z_t^{sp,(E)}$ approaches $1$ as the continuum limit is taken. Note, however, that $z_t^{MB}\leq Z$, with equality being achieved at the continuum limit. Therefore, we expect the localization transition will be marked by a sharper decrease in $z_t^{MB}$ when the continuum limit is taken. Indeed, Fig. \ref{fig: limits}a shows that $z_t^{MB}$ maintains a higher value nearing the transition as the continuum limit is taken. The difference between the $c=4$ curve and the others is visible, but small since the delocalized single-particle states are already almost plane-waves when $c=1$. The effect is less pronounced at lower $V_0/t_0$ because, in that case, the occupied, partially-localized single-particle states can decoupled to plane waves almost exactly. That is, $Z\to1$ as $V_0/t_0\to0$ in this model. Hence, the effect of taking the continuum limit is more pronounced in the delocalized phase only as $V_0/t_0$ is already nearing the transition at $V_0/t_0\approx0.062$.

We now investigate the consequences of taking the continuum limit for $z_x$. Figure \ref{fig: limits}c shows $z^{MB}_x$ for the many-body ground state at $\mu=0$ for different values of $V_0/t_0$. Note that the localization transition is also sharper in the continuum limit for $z^{MB}_x$. In fact, Fig. \ref{fig: limits}c shows that the effect of taking the continuum limit is to pin $z^{MB}_x=0$ throughout the delocalized phase, even near the transition. This makes for a steeper increase of $z^{MB}_x$ at the transition to the partially localized phase, hence the sharper transition. The single-particle $z^{sp}_x$, however, does not change qualitatively as the continuum limit is taken at fixed $V/t$, and its shape is the same as that shown in Fig.\ref{fig: DDM single particle}a (see Appendix \ref{appx: zx_sp} for the effect of the limits in $z_x^{sp}$). This is consistent with our previous discussion explaining why for the DDM the single-particle $z_t^{sp}$ has more bearing on the corresponding many-body indicator $z_{t}^{MB},$ than $z^{sp}_x$ has on $z_{x}^{MB}$ (see Section \ref{DDM subsec 2}). Then the effect of the continuum limit, while not dramatic in the single-particle $z^{sp}_x$ (see Appendix \ref{appx: zx_sp}), still shows clear features in the many-body $z^{MB}_x$.

Interestingly, the effect of the continuum limit on the many-body $z^{MB}_x$ seems to be that it makes $z_x^{MB}\to0$ a better measure of \emph{delocalization}, as it is more strongly pinned to its vanishing(delocalized) value throughout the delocalized phase. That is, in the continuum limit the $z_{x}^{MB}$ indicator more accurately detects delocalized states, e.g., we found that filling even a few delocalized single-particle states in the DDM is enough to send $z^{MB}_x\to0$. We indeed expect $z_x\to0$ for delocalized states from Resta's formula (Eq.\ref{eq: Resta's formula}) when $\lambda\gg L$, but this formula is exact only in the thermodynamic limit, not the continuum limit. Nevertheless, since $[U,T]\neq 0$, the uncertainty relation in Eq. \ref{eq: uncertainty relation 2 zx zt} implies that, while $z_t$ approaches its maximum value $Z$ in the continuum limit, $z_x$ must get closer to $0$. This explains why $z_x$ is pinned at $0$ until immediately before the transition, especially at higher $c$ (see note \footnote{The filling fraction $N_e/L$ at $\mu=0$ is approximately $1/2$ for the DDM in this parameter range, which guarantees that a small $z_t$ implies a large $z_x$ and vice-versa. For small $N_e/L$, such that $U$ and $T$ almost commute, the uncertainty relation places only weak bounds in the relationship between $z_x$ and $z_t$. Therefore, the explanations based on the uncertainty relation are only well-suited for the effect of the thermodynamic limit for $z_t$, and continuum limit for $z_x$, when $[U,T]$ is large enough.}).

With this in mind, let us consider the effect of the thermodynamic limit on both localization measures. In Fig. \ref{fig: limits}d, the many-body $z^{MB}_x$ transitions from zero towards its maximum value at smaller values of $V_0/t_0$ as the thermodynamic limit is taken. That is, the transition happens for smaller $V_0/t_0$ for systems with larger length $L=cL_0$. The maximum value of $z^{MB}_x$ after the transition is not $1$, as discussed previously, as the ground state is only partially localized at $\mu=0$ for this range of $V_0/t_0$. What we observe is that the thermodynamic limit pushes $z^{MB}_x$ to its maximum value faster in the more localized phase, in the same way that the continuum limit pushed $z_t$ to its maximum value in the delocalized phase. We therefore interpret the role of the thermodynamic limit as making $z_x$ a better measure of localization (rather than delocalization). Indeed, Resta's formula shows that the thermodynamic limit makes $z_x$ a better probe for $C^{(X)}_2$, which is smaller for more localized states. Hence, $z_x$ is the largest for the more localized states in the thermodynamic limit, reflecting $z_x=e^{-\frac{1}{2}\left(\frac{2\pi}{L}\right)^2C^{(X)}_2}$.

In comparison, the thermodynamic limit makes the many-body $z^{MB}_t$ decrease more rapidly in the more localized phase, as shown in Fig. \ref{fig: limits}b. This is consistent with the interpretation that the thermodynamic limit also makes $z_t^{MB}\to0$ a better measure for localization (as compared to $z_t^{MB}\to1$ being a better measure of delocalization in the continuum limit). Figure \ref{fig: DDM single particle limits}b shows that in the thermodynamic limit the single-particle $z_t^{sp}$ takes longer to increase toward $1$ for the single-particle states that are expected to be fully delocalized (the upper energy range). Then $z^{sp}_t$ remains smaller in the thermodynamic limit for single-particle states that are already more delocalized, making the many-body $z_t^{\text{MB}}$ also reduce in the thermodynamic limit, as seen in Fig. \ref{fig: limits}b. This is also consistent with the uncertainty relation:  since $z_x$ is the largest possible in the thermodynamic limit for the partially localized states, $z_t$ is suppressed in that limit.

\begin{figure*}
    \centering
    \includegraphics[width=1\linewidth]{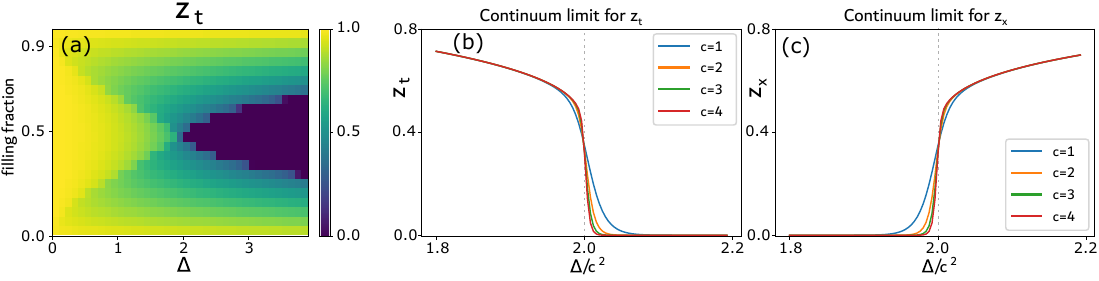}
    \caption{(a) Localization phase diagram of $z_t^{MB}$ for the SSD, with a system of size $N=900$. (b,c) Effect of taking the continuum limit on $z_t$ (b) and $z_x$ (c) across the localization transition, for $L=300$ and fixed filling fraction $0.5$ (half-filling). The transition gets sharper with the continuum limit for both $z_t$ and $z_x$. In particular, $z_t=z_x$ at $\Delta/c^2=2$, the self-dual point where $H_{SSD}$ is the same in position and in momentum space.}
    \label{fig: SSD density}
\end{figure*}

Therefore, our numerics show that the continuum limit makes both measures more accurate for capturing delocalization by making the many-body $z^{MB}_x$ and $z^{MB}_t$
go towards $0$ and $1$, respectively, more sharply as the state becomes delocalized across the phase transition.  Analogously, the thermodynamic limit has the complementary effect of making the many-body $z^{MB}_x$ and $z^{MB}_t$ more accurate measures for capturing localization, i.e., they go towards $1$ and $0$, respectively, more sharply as the state becomes localized across the phase transition. Which of the measures is best to use depends on the application, with $z_x$ being a more natural probe for localization, and $z_t$ for delocalization since they reach their maximum values in those phases. This choice also avoids the subtleties with diverging localization lengths discussed in Ref. \cite{hetenyi-geometric-cumulants}. Indeed, Ref.\cite{hetenyi-geometric-cumulants} claims that the diverging $C_2^{(X)}\to\infty$ of delocalized states, which implies $z_x=0$, is unphysical for finite system sizes. By considering $z_t$ and the finite $C_2^{(K)}$ instead, we bypass this apparent unphysical aspect of the localization measure $z_x$ when used for delocalized states.

Let us now review what we accomplished for the DDM. We have shown numerical evidence that the continuum limit makes $z_t$ a sharper delocalization measure for delocalized states, corroborating our analytic result in Eq. \ref{eq: c2 momentum simple}. Additionally, we used the uncertainty relation between $z_t$ and $z_x$ to explain the effect of the continuum limit on $z_x$ as well. Altogether, we presented an explanation for the role that the continuum and thermodynamic limits play in making $z_{x}$ and $z_t$ better probes of localization (thermodynamic limit) and delocalization (continuum limit).

Although illuminating for our purposes, the DDM was specifically engineered to have clearly defined and distinct continuum and thermodynamic limits. Other models may not enjoy these same attributes. In the next section, we discuss one such example, which we call the Simple Self-Dual Model. There, the relationship between the continuum and thermodynamic limits is more subtle, and we will see how this impacts $z_x$ and $z_t$.

\subsection{Simple Self-Dual Model} \label{sec: SSD}

As another example of a model with a clear continuum limit, we now introduce the Simple Self-Dual model (SSD). This model is special because the Hamiltonian places translations in both position and momentum space on the same footing, which has subtle implications for the interpretations of the limits and their effect on the localization measures. In fact, we claim that the thermodynamic and the continuum limit are the same at the self-dual point of this model.

The SSD Hamiltonian consists of translations in both position and momentum space, and can be written in terms of the modular position and translation operators:
\begin{align}
    H_\text{SSD} &= t(T + T^\dagger) + \frac{\Delta}{2}( U +U^\dagger)\nonumber\\
    &=\sum_{j} t(c^\dagger_j c_{j+1}+h.c.) + \sum_q \frac{\Delta}{2}( c^\dagger_{q}c_{q+1}+h.c.){}\label{eq: HSSD base},
\end{align}
where $j$ is a position index, and $q$, a momentum index. The position-basis description of $H_{SSD}$,
\begin{align}
    H_\text{SSD} &= \sum_{j} t(c^\dagger_j c_{j+1}+h.c.) + \Delta\sum_i \cos{\left(\frac{2\pi}{L} j\right)}c^\dagger_jc_j\label{eq: HSSD},
\end{align}
is a similar to the Aubry-Andre model \cite{aubry_andre_1980,harper_1955} (see Section \ref{sec: AA model}), where the irrational periodicity of the onsite potential is replaced by a potential having period $(2\pi/L)$.

In Fig. \ref{fig: SSD density}a, we show the localization phase diagram, as measured using $z_t.$ When $\Delta = 2t$ in $H_{SSD}$, the coefficients of the terms in the Hamiltonian in both position and momentum space are identical, and at half-filling this is the localization transition point \footnote{ At the point where $t=\Delta/2$, $H_\text{SSD}$ becomes the Harper Hamiltonian \cite{harper_1955}, whose ground state minimizes the uncertainty relation in Eq. \ref{eq: uncertainty relation 2 zx zt}\cite{2008-Massar-Spindel,PRA_uncertainty}}. We find that the phase diagram is symmetric around this transition point. The phase diagram determined using $z_x$ is complementary to that of $z_t$, with low values of $z_x$ where the values of $z_t$ are high.

Now we consider taking the limits of this model, starting with the continuum limit. The onsite potential of this model, $\cos{\left(\frac{2\pi}{L} j\right)}$, is sampled at equally spaced points of the cosine corresponding to the discrete position sites. The entire lattice samples only one period of this potential no matter what system size $L$ is chosen. Hence, in the continuum limit, the total system size $L,$ and hence the potential, are fixed, and the number of sampled points increases as we decrease the inter-site spacing, the lattice constant $a$. 

To instead take the thermodynamic limit, one could try  keeping the lattice spacing fixed and increasing $L$ to $ L_{new}$. The new allowed values of momenta are spaced by $\frac{2\pi}{L_{new}}<\frac{2\pi}{L}$. If we kept the onsite potential unchanged at $\cos{\left(\frac{2\pi}{L} j\right)}$ we see that it mixes only the values of momenta that are $\frac{2\pi}{L}$ apart (assuming $L_{new}=cL$ for $c\in\mathbb{Z})$. Momenta spaced by $\frac{2\pi}{L}$  are no longer adjacent for a system size $L_{new},$  therefore, position and momentum space are no longer in the same footing, since the position-space hopping still happens between adjacent sites. The model would hence be changed to lose the self-duality from before. Since we want the model to remain qualitatively the same in the physical thermodynamic limit, we conclude that this is not the appropriate way to take the limit. 

Alternatively, one can take the thermodynamic limit by taking $\cos{\left(\frac{2\pi}{L} j\right)}\to \cos{\left(\frac{2\pi}{L_{new}} j\right)}$. In this case, there is still only one period of the onsite potential throughout the entire system, and only the number of sampled points increases. That is, the density of sampled points increases in one period of the potential, effectively moving towards the continuum limit. Therefore, the effect of taking the thermodynamic limit like this is the same as that of taking the continuum limit.

With that in mind, we numerically verify that the localization transition shown by $z_t$ becomes sharper around the self-dual point at half-filling when the continuum limit is taken, as shown in Fig. \ref{fig: SSD density}b. The peculiar form of this model, having nearest-neighbor hopping in both momentum and position, makes the transition in $z_x$ mirror that in $z_t$, as shown in Fig. \ref{fig: SSD density}c. The reason is clearest at the transition point where the Hamiltonian is the same in both the momentum and position basis, and so are its eigenstates. Since $z_x$ and $z_t$ are both expectation values of translation in either momentum or position space, their behavior around the transition point should be qualitatively the same in the appropriate directions of the transition,  which explains why the plots in Figs. \ref{fig: SSD density}b,c are mirrored and show $z_x=z_t$ at the transition point.

It is worth noting that, at the transition point where the Hamiltonian is exactly the same in either the momentum or position basis, it makes sense that the thermodynamic and the continuum limit conflate. The thermodynamic limit means increasing the number of sites in the model, at fixed lattice constant $a_0$. This increases the extension of the system in position space. On the other hand, taking the continuum limit increases the extension of the system in momentum space from $2\pi/a_0$ to $2\pi/a$, $a<a_0$, at fixed momentum-space spacing $(\frac{2\pi}{L})$. Therefore, at the self-dual point, increasing the dimension of the Hamiltonian increases the extension of the system in both position and momentum space, enforcing both limits.

\subsection{Models without a continuum limit}\label{sec: no continuum limit}
So far we have discussed only models that have a clear continuum limit. However, it is often the case that models naturally defined on a lattice do not have a well-defined or obvious continuum realization. Four our a last examples, we consider whether $z_t$ still has any meaning as a localization measure in models that lack a clear continuum limit.

We consider two models with a clear localization transition to probe the behavior of $z_t$ when the continuum limit is not well-defined. The first is the disordered Random Dimer model (RDM)\cite{philip_RDM_1990}, and the second, the quasiperiodic Aubry-Andre model {\cite{aubry_andre_1980,harper_1955}}. We numerically show that $z_t$ is not as clear a probe (at least by eye) of the localization transitions as $z_x$ in either model, and point out the different mechanisms for this behavior in each model. Nevertheless, $z_t$ is still sensitive to the location of the localization transition in both models. We demonstrate that this sensitivity can be empirically seen from the first derivative of the momentum localization length $\lambda_k$.

\subsubsection{Random Dimer Model} \label{sec: RDM}
The Random Dimer Hamiltonian
\begin{equation}
    H_\text{RDM} = \sum_{j}\epsilon_jc^\dagger_jc_j + \sum_j t(c^\dagger_jc_{j+1}+h.c.),
\end{equation}
has a disordered on-site potential $\epsilon_j$, which may take two possible values $\epsilon_a$ or $\epsilon_b$ at each site $j$ \cite{philip_RDM_1990,taylor-dimer-momentum-spectrum}. \footnote{In momentum space, $H_\text{RDM}$ is similar to our deterministic dimer model's Hamiltonian. The main difference is that $V$ must take random imaginary phases to account for the randomness of the potential. The exact expression is given in Ref. \cite{taylor-dimer-momentum-spectrum}.} The on-site energy values are randomly placed with the constraint that, whenever $\epsilon_a$ is randomly placed on site $j$, it also fixes $\epsilon_{j+1}=\epsilon_a$. This scheme for randomly placing the onsite potential in the 1D lattice does not allow a natural interpolation to the continuum limit by reducing the lattice constant as we did for the other models.

\begin{figure}
    \centering
    \includegraphics[width=1.1\linewidth]{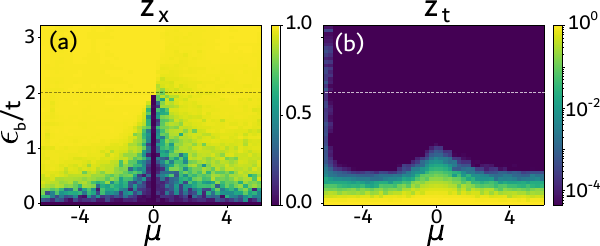}
    \caption{Localization phase diagrams of the RDM for $z_x^{MB}$ (a) and $z_t^{MB}$ (b). (a) The delocalized phase, having $z_x^{MB}\to0$ is in dark blue, and the localized phase, having $z_x\to1$, is in bright yellow. At $\mu=0$ (the resonant energy), the localization transition happens at $\epsilon_b/t=2$, which is clear (by eye) only in (a). (b) The bright yellow region indicates $z_t\to1$, and the dark blue region indicates $z_t\to0$. As discussed in the main text, $z_t^{MB}$ has a weaker indication of the localization properties of the RDM, since this model does not have a clear interpolation to the continuum limit that would sharpen the features.}
    \label{fig:RDM zx density}
\end{figure}

When $|\epsilon_b-\epsilon_a|<2t$, $\mathcal{O}(\sqrt{L})$ of the single-particle states that have energy in the vicinity of the resonant energy $\epsilon_a$ become delocalized, while the other eigenstates remain localized in position space. Hence, for large $L$, most of the the single-particle eigenstates are localized, except a small number of states in the vicinity of $\epsilon_a.$ Thus, we expect  the single-particle $z_x$ may be used to approximately infer the many-body localization properties, as discussed in Section \ref{subsec: DDM}, while the single-particle $z_t^{sp}$ will not straightforwardly reflect the localization of the many-body ground state. We will show numerical evidence that, indeed, the many-body $z_x^{MB}$ follows the single-particle localization properties described here, e.g., Fig. \ref{fig:RDM zx density}(a).

Intuitively, for a fixed disorder strength $|\epsilon_b-\epsilon_a|<2t$, the many-body ground state is localized while the chemical potential is below the energy of the first delocalized single-particle eigenstate, or above the energy of the last delocalized single-particle state. At $\mu\approx \epsilon_a$, in contrast, the ground state is delocalized. The delocalization at  $\mu\approx \epsilon_a$ can be seen in Fig. \ref{fig:RDM zx density}(a), where we plot the many-body $z_x^{MB}$ as a function of disorder strength and chemical potential, and take $\epsilon_a=0$.  Note that there is no single-particle delocalization for disorder strengths $|\epsilon_b-\epsilon_a|>2t$, which explains why Fig. \ref{fig:RDM zx density}(a) shows that the state is localized $(z_x^{MB}=1)$ for any $\mu$ at that disorder stength.

\begin{figure}
    \centering
    \includegraphics[width=0.6\linewidth]{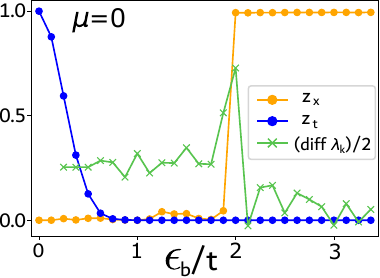}
    \caption{Detail of the localization transition for the many-body ground state at $\mu=0$. The orange curve shows that $z_x$ accurately marks the localization transition, while the blue curve shows that $z_t$ dramatically decays before the transition point $\epsilon_b/t=2$. However, the transition information is still contained in the $z_t$ curve as the green line marked with 'x' shows that the finite difference (discrete derivative) of $\lambda_k$ computed from $z_t$ peaks exactly at $\epsilon_b/t=2$, accurately marking the transition point.}
    \label{fig: RDM zx zt diff}
\end{figure}

While there is no clear interpolation between the lattice and the continuum limit for the RDM to make $z_t$ a sharper localization measure, we numerically verify that $z_t$, calculated for the many-body ground state, still picks up the location of the transition, albeit with a weaker trend. 
\begin{figure}
    \centering
    \includegraphics[width=1\linewidth]{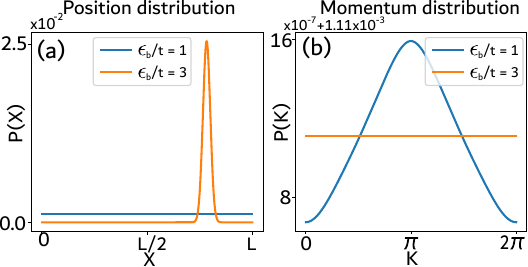}
    \caption{Many-body position (a) and momentum (b) distributions of the ground state of the RDM deep in the delocalized (blue) and localized (orange) phases. The scale of the blue peak in momentum space (b) is much smaller than the orange peak in position space (a), reflecting the fact that $z_t$ decays too quickly and is not a good localization measure for the RDM even in the delocalized phase. }
    \label{fig:RDM distributions}
\end{figure}
\begin{figure*}
        \centering
    \includegraphics[width=0.86\linewidth]{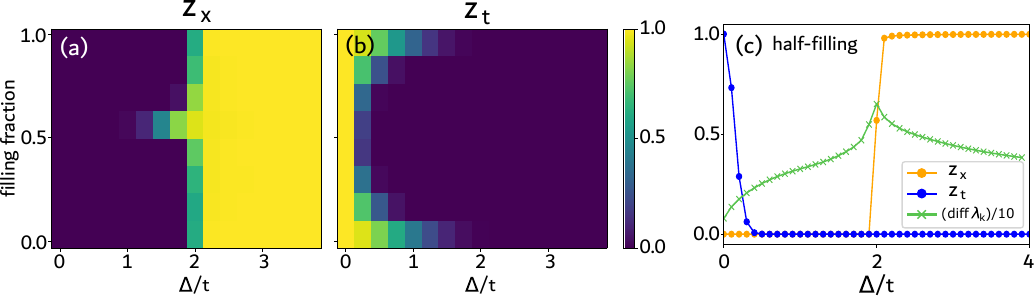}
    \caption{(a,b) Localization phase diagrams for the Aubry-Andre model. The measure $z_x$ (a) accurately detects that the ground state of the AA model is completely delocalized at any filling for $\Delta/t<2$. In contrast, $z_t$ (b) decays quickly as $\Delta/t$ increases at any filling fraction, with no clear mark of the transition at $\Delta/t=2$. (c) Comparison between $z_x$ (orange line), $z_t$ (blue line), and the finite difference of $\lambda_k$ (green line) at half-filling across the localization transition. The finite derivative of $\lambda_k$ peaks at the correct transition point. }
    \label{fig: AA combined}
\end{figure*}
The main pitfall of $z_t$ as a localization measure for the RDM is that it decays rapidly well-before the actual localization transition. This can be seen in Fig. \ref{fig:RDM zx density}(b), where $z_t$ approaches $0$ as a function of disorder strength well-before before the transition point in the vicinity of the delocalized single-particle eigenstates at $\mu=\epsilon_a=0$. A comparison between $z_x$ and $z_t$ at $\mu=0$ for different disorder strengths is shown in Fig. \ref{fig: RDM zx zt diff}, which corresponds to fixing $\mu=0$ in Fig.\ref{fig:RDM zx density}. Note that the real localization transition,  clearly captured by $z_x$, occurs at $\epsilon_b/t=2$, while $z_t$ falls towards zero rapidly as $\epsilon_b$ increases away from the clean limit.

The early decline of $z_t$ before the critical $\epsilon_b=2t$ obscures the determination of the critical point from $z_t$ alone, but we can compute the momentum distribution to gather extra information. We find that the sharp decrease in $z_t$ corresponds to a peaked momentum distribution where the peak amplitude strongly decreases as disorder strength increases. This can be seen in Fig. \ref{fig:RDM distributions}, where the many-body position and momentum distributions of the ground state of the RDM at half-filling are plotted in the localized and in the delocalized phase for comparison. The plots show that, in the delocalized phase, the momentum distribution peak is wider (as a fraction of $2\pi/a$) than the position distribution peak in the localized phase (as a fraction of $L$). This explains why we see $z_x\to1$ in the localized phase while $z_t\ll 1$ in the delocalized phase. This is not inconsistent with our result that $z_t\to1$ in completely delocalized states, since we require the continuum limit to be taken to even define the notion of complete delocalization. 

Since the peak in the momentum distribution of the RDM ground state disappears exactly at $\epsilon_b=2t$, the physical transition point, we might expect that $z_t$ is still sensitive to the localization transition despite its rapid decrease. In fact, we find that the transition is marked by the point where the momentum localization length, 
\begin{equation}
    \lambda_k=\sqrt{C_2^{(K)}}=\sqrt{-(2/a^2)\log z_t},
\end{equation} has its steepest increase. To demonstrate, in Fig. \ref{fig: RDM zx zt diff} we find that the transition point $\epsilon_b=2t$ coincides with a maximum in the finite difference of the momentum localization length (green line). In conclusion, the momentum distribution and $z_t$ remain sensitive to the localization transition even in lattice models where the continuum limit is not well defined. However, $z_t$ itself may not have as much overt contrast between phases as $z_x$ when magnitudes are directly compared.

\subsubsection{Aubry-Andre model}\label{sec: AA model}
The final model we consider is the Aubre-Andre model. This model is a quasiperiodic model with  Hamiltonian
\begin{equation}
    H_\text{AA} = \sum_{j} t(c^\dagger_j c_{j+1}+h.c.) + \Delta\sum_j \cos{\left(2\pi\beta j\right)}c^\dagger_jc_j\label{eq: Aubry Andre},
\end{equation}
where $\beta$ is incommensurate with the underlying lattice periodicity, and is taken to be the irrational Golden ratio $\varphi = (\sqrt{5}-1)/2$. The model is self-dual at $\Delta = 2{t}$,  where the Hamiltonian has the same form when written in either the position or a generalized momentum basis.\cite{hobbyhorse} These position and generalized momentum bases are related by a Fourier Transform where the exponential factors take the form $e^{i2\pi\beta k j}$, for $j$ a position index \cite{hobbyhorse}. At the point $\Delta=2t$, the system undergoes a transition where all the eigenstates switch from delocalized to localized as $\Delta$ increases \cite{aubry_andre_1980,harper_1955,hobbyhorse}.

The Aubry-Andre model has a well-defined thermodynamic limit, and the scaling of the localization length with the system size has been investigated in depth in Ref. \cite{hetenyi-generating_reference}. However, a continuum generalization of the Aubry-Andre model which exactly preserves all of its localization properties is not known, to the best of our knowledge \cite{schreiber_exp_bichromatic_2015} \footnote{In experiments having incommensurate potentials, where the tight-binding limit is not a good approximation of the system, there are single-particle mobility edges which are not explained by the AA model \cite{schreiber_exp_bichromatic_2015}. There exists a 1D bichromatic incommensurate model which appropriately accounts for the continuum degrees of freedom to explain the effects seen in experiment which differ from the AA model, as shown in Ref. \cite{sarma_bichromatic_2017}. The Aubry-Andre model is a deep tight-binding limit of this continuum model, but they show different localization properties, as the continuum model has an intermediate localization phase which is absent from the AA model.}. Additionally, the need to employ a commensurate approximation scheme of the onsite potential to numerically simulate the model\cite{goncalves_comensurate_2022}\footnote{In order to numerically simulate a periodic 1D lattice with an incommensurate potential, we employ a commensurate approximation for the Golden ratio, using $\beta = n_1/n_2$, where $n_1<n_2$ are adjacent integers in the Fibonacci sequence. The ratio $n_1/n_2$ approaches the irrational $\varphi$ for large numbers in the sequence. The periodicity of the chain is ensured by setting the total number of lattice sites $N$ to be integer multiples of $n_2$.} also obstructs a numerical interpolation to the continuum limit such as the one we employ for the DDM. Therefore, the Aubry-Andre model is an example of a model having a clear localization transition that does not have a well-defined continuum limit with the same localization properties.

Having seen the limitations of using $z_t$ as a localization measure for the RDM in the previous section, it is natural to ask whether the quasiperiodicity of the Aubry-Andre model makes $z_t$ a better measure in this model than in the RDM. Here, we numerically demonstrate that this is not the case, despite the momentum distribution showing different features in the delocalized phase of the AA model.

To demonstrate the claims above, we first show in Fig. \ref{fig: AA combined} (a,b) the many-body $z_x$ and $z_t$ of the AA model as a function of $\Delta/t$ at different filling. The $z_x$ plot shows a clear localization transition at $\Delta/t$ for any filling, reflecting the fact that all the eigenstates undergo the transition in this model (there is no mobility edge as a function of filling). The $z_t$ plot, however, does not show such a clear transition, regardless of the color grading chosen for the plot. Indeed, Fig. \ref{fig: AA combined}b shows that $z_t$ decays to zero before the self-dual point, similarly to how $z_t$ decayed before $\epsilon_b=2t$ in the RDM in Fig. \ref{fig:RDM zx density}b.

\begin{figure}
        \centering
        \includegraphics[width=1\linewidth]{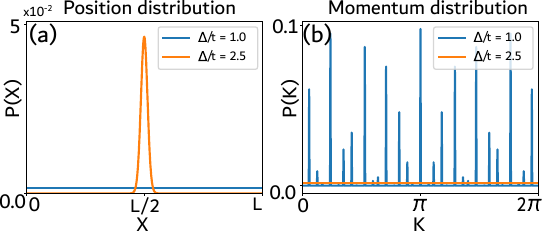}
        \caption{Position (a) and momentum (b) distributions of the ground state of the Aubry-Andre model at half-filling. The blue curve corresponds to $\Delta/t =1$, in the delocalized phase, and the orange curve corresponds to $\Delta/t=2.5$, in the localized phase. The blue (delocalized) state has a quasiperiodic momentum distribution with sharp peaks (b), which explains why $z_t$ decays quickly even in the delocalized phase and is not a good delocalization measure for the AA model.}
        \label{fig:AA distributions}
    \end{figure}

The behavior of $z_x$ and $z_t$ across the transition is compared in more detail for a half-filled system in Fig. \ref{fig: AA combined}c. Note that $z_x=0$ for $\Delta/t<2$, and $z_x=1$ for $\Delta/t>2$, accurately marking the localization transition at $\Delta/t=2$. In contrast, $z_t$ decreases sharply even before the transition. This sharp decrease of $z_t$ may be understood from the momentum distribution of the Aubry-Andre model, shown in Fig. \ref{fig:AA distributions},  alongside the position distribution for different values of $\Delta/t$. The position distribution (Fig. \ref{fig:AA distributions}a) is flat in the delocalized phase, and shows a localized peak in the localized phase, as expected. The momentum distribution (Fig. \ref{fig:AA distributions}b), however, is not completely localized in momentum space even when $\Delta/t<2$. In fact, the quasiperiodicity of the model generates a momentum distribution exhibiting a pattern of narrow peaks spread throughout momentum space, as shown in Fig. \ref{fig:AA distributions}b. Since the second cumulant of this pattern of peaks in the momentum distribution is not vanishing, we find $z_t<1$ even when $\Delta/t<2$, that is, in the delocalized phase. As $\Delta/t$ increases to approach the transition point, the number of peaks grows, causing their magnitude to decrease and the momentum distribution to approach a more uniform distribution. This explains the rapid decline in $z_t$ even before the transition. However, only when $\Delta/t\geq2$ do the peaks truly disappear. Hence, we can still see the transition directly in the momentum distribution.

As before, we find that we can extract the transition point from $z_t$ itself by computing the maximum in the finite differences of $\lambda_k$, shown in Fig. \ref{fig: AA combined}c. Therefore, although $z_t$ itself is not a high-contrast indicator of the localization transition of the AA model, $\lambda_k$ extracted from $z_t$ is still sensitive to the correct transition point, similarly to our findings for the RDM model. However, note that the mechanisms for $z_t$ to decay before the physical transition are different for the AA and the RDM: the former is explained by the quasiperiodicity of the model generating many sharp peaks throughout the momentum distribution in the delocalized phase, while the latter is explained by the widening of a single peak in the momentum distribution of the ground state. It would be interesting for future work to identify if there are other mechanisms for localization transitions that can be visualized in the momentum distribution beyond these two.

\section{Flux Insertion} \label{sec: flux insertion}

Having studied how the magnitude of the expectation value of the translation operator $z_t = |\langle \Psi|\hat{T}|\Psi\rangle|$ captures the localization properties of states that break translation symmetry, we now turn to studying how the phase $\arg\langle \Psi|\hat{T}|\Psi\rangle$ may do the same. 

In translation-symmetric states, $\langle\Psi|\hat{T}|\Psi\rangle = e^{iaK_0}$, where $K_0$ is the total momentum of the state $|\Psi\rangle$. As reviewed in Section \ref{sec: background}, Ref. \cite{nonzero_momentum} showed that only LRE states may have the value of $K_0$, and hence the phase $\arg\langle\Psi|\hat{T}|\Psi\rangle $, changed by inserting flux into the system. Additionally, the only values of $K_0$ allowed for SRE states are the trivial ones, $K_0=0$ for bosons, or $K_0\in\{0,\pi\}$ for fermions. Therefore, SRE states cannot achieve any nontrivial value of momentum via flux insertion in translation-symmetric systems.

Here we answer the question: does inserting flux also leave the momentum distribution of a SRE state invariant when translation symmetry is broken? As mentioned in the Introduction, localized states are a good proxy for SRE states in 1D periodic systems. We demonstrate using three examples that inserting flux shifts $\arg\langle\Psi|T|\Psi\rangle$ and the momentum distribution in states that break translation symmetry \textit{regardless of their localization properties}. However, the key point is that the final distribution is indistinguishable from the original one \emph{only when the state is completely localized in position space and the momentum distribution is flat} (has $C_{2}^{(K)}\to\infty$). This apparent invariance under flux insertion may be used to distinguish between localized/SRE and delocalized/LRE states even in the absence of translation symmetry. The results are summarized schematically in Fig. \ref{fig: distribution schematic}.

We provide three examples to illustrate how flux insertion affects the momentum distribution of states that break translation symmetry. The first example is a gaussian-localized state in position space, which allows for a direct analytic calculation of the expectation value of the translation operator and its phase. The gaussian state example is also useful because a gaussian distribution interpolates between a completely uniform distribution and a delta function, as its standard deviation is tuned. We further use the gaussian state calculations to show that the completely filled lowest Landau level in a Landau gauge (which breaks ordinary translation symmetry while preserving magnetic translations) can be seen to  be insensitive to flux in the localized direction, and sensitive to flux in the delocalized direction. Finally, we provide numerical evidence that the momentum distribution of the many-body ground state of the RDM is sensitive to the insertion of flux only in the delocalized phase. Here, being ``sensitive" to flux insertion means that the momentum distribution after flux insertion is distinguishable from the momentum distribution prior to flux insertion.

\subsection{Gaussian states} \label{sec: Gaussian states}
The behavior of the momentum distribution under flux insertion can be easily analytically studied for Gaussian states, which have the same form in the position and in the momentum basis. In the latter, it reads
\begin{equation} \label{eq: Gaussian state momentum}
    \Psi(k) = \frac{1}{\sqrt{\sqrt{\pi}\sigma_k}}e^{-\frac{(k-k_0)^2}{2\sigma^2_k}}
\end{equation}
for some momentum $k$, where $\sigma_k$ is the standard deviation of the momentum distribution. When the spread in momentum is of the order $\sigma_k\geq 2\pi/a$, the state is localized in position space, since $\sigma_x=1/\sigma_k$ for Gaussian states. The localization lengths $\lambda_{x,k}$ are related to $\sigma_x$ and $\sigma_k$ via $\lambda_x=\frac{\sigma_x}{\sqrt{2}}=\frac{1}{\sqrt2\sigma_k}=\frac{1}{2\lambda_k}$.

Before inserting flux, the overlap between the Gaussian state in Eq. \ref{eq: Gaussian state momentum} and a plane wave with momentum $k'$, $|\phi_{k'}\rangle$, is given by $\langle \phi_{k'}|\Psi\rangle = ({\sqrt{\pi}\sigma_k})^{-\frac{1}{2}}e^{-\frac{(k'-k_0)^2}{2\sigma_k^2}}$. Once flux is inserted, the momentum of each plane wave is shifted by a constant value $A$, such that $k'\to k'+A$. Then the new overlap is $\langle \phi_{k'}|\Psi\rangle =({\sqrt{\pi}\sigma_k})^{-\frac{1}{2}} e^{-\frac{(k'+A-k_0)^2}{2\sigma_k^2}}$.

We now investigate if $\arg\langle\Psi|T|\Psi\rangle$ may distinguish between localized and delocalized Gaussian states. Setting $k_0=0$ without loss of generality, the translation expectation value with flux insertion can be written as
\begin{align}
    \langle \Psi|T_1|\Psi\rangle &= \frac{1}{\sqrt{\pi}\sigma_k}\int dk e^{-\frac{(k+A)^2}{2\sigma_k^2}}e^{iak}e^{-\frac{(k+A)^2}{2\sigma_k^2}}\\
    &= \frac{1}{\sqrt{\pi}\sigma_k}\int dk e^{iak}  e^{-\frac{(k+A)^2}{\sigma_k^2}}\nonumber\\
    &= \frac{1}{\sqrt{\pi}\sigma_k}\int dk' e^{ia(k'+A)}  e^{-\frac{(k')^2}{\sigma_k^2}}\nonumber\\
    &= \frac{1}{\sqrt{\pi}\sigma_k}e^{iaA}\int dk e^{iak}  e^{-\frac{k^2}{\sigma_k^2}}=e^{iaA}e^{-\frac{\sigma^2_ka^2}{4}}.\label{eq: flux sensitivity}
\end{align}
Note that the integral in the last line is the exact same as without the flux. In fact, the flux-dependent phase factor appears as a multiplicative constant that is independent of $\sigma_k$. Therefore, any change in the phase $\arg \langle\Psi|T|\Psi\rangle$ from flux insertion is \textit{independent of the localization properties} of the state. 

However, it is still possible to distinguish localized from delocalized states from how their momentum distributions change under flux insertion. Note that, when the state is delocalized in position space and localized in momentum space, corresponding to $\sigma_ka\to0$, the magnitude $z_t=|\langle T\rangle|\to1$ is finite and the phase of $\langle T\rangle$ is well-defined, showing the expected flux sensitivity in Eq. \ref{eq: flux sensitivity}. On the other hand, in the limit where $\sigma_ka \gg 1$, the state is localized and the magnitude vanishes $|\langle T\rangle|\to 0$, which makes any phase ill-defined. That is, the phase is not sensitive to flux simply because \textit{it is not well-defined in localized states} when translation symmetry is absent.

The ill-definition of the phase, and hence the flux insensitivity of the momentum distribution, may also be understood by noting that $\sigma_k\gg1/a$ implies a completely flat momentum probability distribution, with $z_t\to0$. Therefore, any shifts to this distribution can take it only to itself, thus having no effect on the state. That is, any phase $e^{iaA}$ of $\langle T \rangle$ has no physical consequence when the state is localized in position space with a flat momentum distribution. We numerically illustrate this situation with the RDM example in Section \ref{sec: RDM flux insertion}.

Now we turn to the intermediate case where  $\sigma_ka$ is of order $1$, as for a nonzero finite $a$ and $\sigma_k$. Then Eq. \ref{eq: flux sensitivity} gives that $z_t=|\langle T\rangle|>0$ is nonzero, and the phase is still technically well-defined \textit{and sensitive to flux}, even though $\sigma$ may extend through the entire momentum space, $\sigma_k \approx 2\pi/a$. Note that $\sqrt{2}\lambda_x=\sigma_x= 1/\sigma_k \approx \frac{a}{2\pi}$, so Resta's formula in Eq. \ref{eq: Resta's formula} gives $z_x\approx e^{-\frac{1}{4}\left(\frac{a}{L}\right)^2}$. Therefore, states having $\sigma_k \approx 2\pi/a$ are localized in the sense that $z_x\to1$ in the thermodynamic limit, while simultaneously being sensitive to flux.
While the momentum distribution is almost flat, it is not completely uniform, and inserting flux shifts the distribution to a similar, but tractably distinct one.

The example above for $\sigma_k\approx2\pi/a$  shows a possible feature of localized states that break translation symmetry which is absent in localized translation-symmetric states in 1D. The latter are always insensitive to flux insertion, as shown in Ref. \cite{nonzero_momentum}, while the former may be simultaneously localized and sensitive to flux as long as the momentum distribution is not completely flat, as in the case of $\sigma_k\approx2\pi/a$ for finite $a$. Note that the continuum limit, however, $\sigma_k\approx2\pi/a\to \infty$, and the distribution is completely uniform again. Therefore, the intermediate case $\sigma_k\approx2\pi/a$  is only both sensitive to flux insertion and localized away from the continuum limit.

\subsection{Lowest Landau Level}\label{sec: LLL}
We will now demonstrate that the Lowest Landau Level (LLL) in a Landau gauge provides a clear example of how localized and delocalized states respond differently to flux insertion. We preface this example by recognizing that while the localization properties of the single-particle states of a filled LLL are gauge dependent, one localization property that is clearly depicted with this choice of gauge is that the single-particle orbitals comprising the LLL cannot be simultaneously localized in both spatial directions due to the nontrivial Chern number of the system. Even though the LLL preserves magnetic translations, it breaks the ordinary translation symmetry, i.e., the Hamiltonian does not commute with the conventional translation operators considered so far. Hence, by applying our methods will see that the LLL in the Landau gauge remains a useful toy model to illustrate how localization and delocalization influence a state's sensitivity to flux insertion. Moreover, while this is a 2D system, the $x$ and $y$-directions are independent, in the sense that the single-particle wavefunctions factorize into $x$ and a $y$-dependent parts. Therefore, we can treat the LLL as a quasi-1D system for localization purposes.

The LLL single-particle wavefunctions in a Landau gauge $A_y= B x$ have the form
\begin{equation}
    \psi_{k_y}(x,y) =\frac{1}{\sqrt{ \mathcal{N}}} e^{ik_yy}e^{-{(x-k_yl^2_B)^2}/{2l^2_B}},\label{eq: LLL}
\end{equation}
where $\mathcal{N}$ is a normalization constant \cite{tong2016lecturesquantumhalleffect}. These wavefunctions are (delocalized) plane waves in the $y$-direction, and labeled by $k_y$. In the $x$-direction, they are localized gaussians, centered at $x_c(k_y)=k_yl_B^2$, where $l_B=\sqrt{\hbar/eB}$ is the magnetic length. The ground state of the completely filled LLL, denoted $|\Psi_{LLL}\rangle$, can be written as a Slater determinant of all single-particle states $|\psi_{k_y}\rangle$. Therefore, Eq. \ref{eq: overlaps general} can still be used to calculate $\langle \Psi_{LLL}|T|\Psi_{LLL}\rangle$ and investigate the effect on this expectation value of inserting flux in different directions.

We now investigate the effect of inserting flux to shift $k_x\to k_x+A$. Since the $x$-direction is localized, we expect $z_t=|\langle T_x\rangle|$ to vanish and the momentum distribution in the $x$-direction to be insensitive to flux insertion. Indeed, after doing the Fourier transform to write 
\begin{equation}
    \tilde\psi_{k_y}(k_x,y)= \frac{\sqrt{2\pi}l_B}{\sqrt{\mathcal{N}}} e^{ik_y(y+k_xl_B^2)} e^{\frac{-k_x^2l_B^2}{2}},
\end{equation}
and inserting flux to shift $k_x\to k_x+A$, we obtain
\begin{align}
S_{ij}&=\langle\psi_{k_y^{(i)}}|T_x|\psi_{k_y^{(j)}}\rangle\nonumber \\
&=\frac{1}{\mathcal N}\int dy\ \nonumber e^{i(k_y^{(j)}-k_y^{(i)})y}\\& \ \ \ \ \ \ \ \ \times \int dk_x e^{i(k_x-A)l_B^2(k_y^{(j)}-k_y^{(i)})}e^{iak_x}e^{-{(k_x-A)^2l_B^2}}\nonumber \\
&=\frac{1}{\mathcal N}\delta_{ij}\int dk_x e^{iak_x}e^{-{(k_x-A)^2l_B^2}}\nonumber\\
&=\delta_{ij}e^{iaA}e^{\frac{-a^2}{4l_B^2}}.
\end{align}
Since the matrix $S$ is diagonal, 
\begin{align}
        \langle T_x\rangle &= \det S=\prod^{N_e}_j e^{iaA}e^{-\frac{a^2}{4l_B^2}},
\end{align}
where $N_e=L_xL_y/2\pi l_B^2$ is the total number of electrons in the completely filled LLL, and $L_xL_y$ is the area of the system. Then the expectation value of the translation operator in the $x$-direction is
\begin{equation}
    \langle T_x\rangle =e^{iaA\frac{L_xL_y}{2\pi l_B^2}}e^{-\frac{a^2L_xL_y}{8\pi l_B^4}}.
\end{equation}

Note that the magnitude $|\langle T_x\rangle|$ decays exponentially with the area of the system, vanishing in the thermodynamic limit. Another way to see this is to recall that the long-wavelength limit corresponds to the LLL gap increasing, and sending $l_B^2\to0$. Note that the length $a$ by which the operator $T_x$ translates the system is arbitrary, since the continuum limit has already been implicitly taken when writing the LLL wavefunctions. Then the numerator in the magnitude $|\langle T_x\rangle|$ stays fixed while $l^2_B\to0$ in the IR limit, such that $|\langle T_x\rangle|\to 0$. Therefore, any shift in $\arg \langle T_x\rangle = e^{iaA\frac{L_xL_y}{2\pi l_B^2}}$ is not detectable in the momentum distribution in the $x$-direction, and we consider it to be insensitive to flux.

When flux is inserted to shift $k_y\to k_y+A$ instead, the effect on the momentum distribution is different. Upon inserting flux, the single-particle wavefunctions become
\begin{equation}
\psi_{k_y}(x,y) =\frac{1}{\sqrt{\mathcal N}}e^{i(k_y-A)y}e^{-{[x-(k_y-A)l^2_B]^2}/{2l^2_B}},\label{eq: LLL with flux ky}.
\end{equation}
The effect of flux insertion on the plane-wave part is simply of adding a phase, and the $x$-direction is only affected by the center of the gaussians shifting by $Al_B^2$. The elements of the $S$ matrix are given by
\begin{align}
    S_{ij}&=\langle \psi_{k_y^{(i)}}|T_y|\psi_{k_y^{(j)}}\rangle\\
    &= \delta_{i,j}e^{i(k_y^{(j)}-A)a}.
\end{align}
Then we find that the expectation value of the translation operator in the $y$-direction is
\begin{equation}
    \langle T_y\rangle=\det S=\prod_j^{N_e}e^{i(k_y^{(j)}-A)a}=e^{-iaA\frac{L_xL_y}{2\pi l_B^2}},
\end{equation}
a pure phase which is always well-defined. Therefore, we confirm our expectation that the delocalized direction of the LLL, the $y$-direction, is sensitive to flux insertion, while the localized $x$-direction is not. This illustrates that in systems where translation symmetry is broken, localized states can be insensitive to flux because their momentum distribution is completely uniform ($|\langle T\rangle|\to0$), which makes the phase $\arg \langle T\rangle $ ill-defined. 

One might wonder what the expectation value of the magnetic translation operator  reveals about the localization of the states. However, it is the regular translation operator which is the generating function of the momentum distribution, so we consider only that operator here. Additionally, the Landau levels preserve magnetic translations, and we are interested in states that are not eigenstates of the translation symmetry under consideration. Further studies of the magnetic translation operator and localization/entanglement, perhaps for Landau levels in inhomogeneous fields, or in moir\'e systems are left for future work. Another future direction is to relate the physical, gauge-independent localization properties of the LLL to the expectation values of both regular and magnetic translation operators.

In the next subsection, we numerically illustrate the momentum distribution shifting in the presence of flux for delocalized states, while being invariant for localized states.

\subsection{Random Dimer ground state with flux} \label{sec: RDM flux insertion}

We numerically illustrate the effects of inserting flux into a system that breaks translation symmetry by showing how the momentum distribution of the ground state of the RDM at half-filling changes as flux is inserted.

\begin{figure*}
        \centering
    \includegraphics[width=\linewidth]{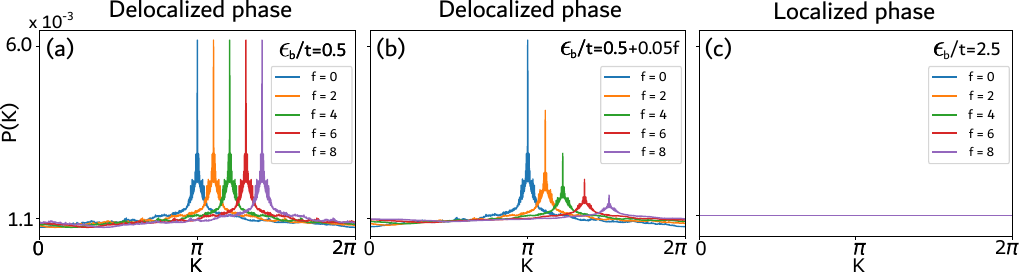}
    \caption{Momentum distributions of the ground state of the RDM at $\mu=0$, for different values of flux insertion. (a) At $\epsilon_b/t=0.5$, the state is in the delocalized phase, and inserting flux shifts the momentum distribution. (b) Increasing $\epsilon_b/t$ quickly decreases the peak in momentum space, but the shift with flux insertion is still detectable as long as the distribution is not completely flat ($z_t\to0$). The peak only disappears after $\epsilon_b/t=2$. (c) The momentum distribution at $\epsilon_b/t = 2.5$, in the localized phase, is completely flat. The distributions for different values of flux are indistinguishable from each other.}
    \label{fig: FLUX RDM}
\end{figure*}

We obtain the momentum distributions for the many-body ground states by doing a discrete Fourier transform of the generating functions, which are expectation values of translation operators by different numbers of lattice sites:
\begin{equation}
    P(K) = \sum_{n=0}^{N-1}e^{iK(an)}f_n = \sum_{n=0}^{N-1}e^{iK(an)}\langle T_n\rangle,
\end{equation}
where $N=L/a$ for $L$ the system size, and the expectation value $\langle T_n\rangle$ is calculated for the many-body ground state from the filled single-particle eigenstates as in Eq. \ref{eq: overlaps general}.

We show the results of the flux insertion for different values of $|\epsilon_b-\epsilon_a|$
in Fig. \ref{fig: FLUX RDM}. Each subfigure of Fig. \ref{fig: FLUX RDM} corresponds to a different value of $\epsilon_b$ at fixed $\epsilon_a=0.0$.
Figure \ref{fig: FLUX RDM}a shows the momentum distribution for $|\epsilon_b-\epsilon_a|<2t$ as flux is inserted. Note that the distributions show a clear peak, which can be tracked to move as flux is inserted. As discussed in Section \ref{sec: RDM}, increasing the disorder strength in the delocalized phase of the RDM decreases the amplitude of the peaks, which also become more spread out. This effect is shown in Fig. \ref{fig: FLUX RDM}b , where the peaks of the momentum distribution are seen to decrease with increasing $\epsilon_b$. While the amplitudes of the peaks decrease, the peak structure is still present until the transition at $|\epsilon_b-\epsilon_a|=2t$, so the shift in the momentum distribution as flux is inserted can be tracked throughout the delocalized phase. In contrast, for $|\epsilon_b-\epsilon_a|>2t$ the peaks disappear and the momentum distribution is completely uniform/flat, as shown in Fig.\ref{fig: FLUX RDM}c. Note that the distributions corresponding to different values of flux in this delocalized phase are indistinguishable from one another by eye in the Figure. Even if they do shift with flux insertion, that shift is essentially undetectable, and there is no physical sense in which the state is sensitive to flux in this localized phase.

This example corroborates our claims that the momentum distribution in localized states is insensitive to flux, and that the flux-dependent phase  $\arg \langle T\rangle$ is not physical when $|\langle T\rangle|=0$. Therefore, flux sensitivity may still be used to distinguish localized and delocalized states in systems that break translation symmetry, but not the phase $\arg\langle T\rangle$ directly, since it is not always well-defined.

\section{Conclusions}\label{sec: conclusion}
We have proven an analogous relation to Resta's formula in momentum space, showing that the magnitude of the expectation value of the translation operator $z_t$ approaches $1$ for delocalized states in the continuum limit. Since the localization properties of 1D systems with periodic boundary conditions can be mapped to their entanglement properties, with delocalized states being LRE, our result provides a simple way to characterize the entanglement of states which break translation symmetry. Importantly, $z_t$ alone is insufficient to determine whether \textit{translation-symmetric} states are SRE or LRE, and the phase $\arg\langle T\rangle $ can be used instead, as shown in  Ref. \cite{nonzero_momentum}.

For states which break translation symmetry, we further show that the phase $\arg\langle T\rangle$ may distinguish between localized (SRE) and delocalized (LRE) states in the sense that it is well-defined only for delocalized states. This is a consequence of $z_t\to0$ in localized states, and is reflected on the momentum distribution being completely flat. A flat momentum distribution may only be shifted by flux insertion to another flat momentum distribution, indistinguishable from the first. In this sense, only delocalized states are sensitive to flux. Therefore, we have characterized how the magnitude and the phase of the expectation value of the momentum operator may be used to study the localization of states which break translation symmetry in one-dimensional systems with periodic boundary conditions, and their entanglement properties by proxy, extending the results of Ref. \cite{nonzero_momentum}.

We validated our results with several numerical examples. In particular, we introduced two new models, the Deterministic Dimer model and the Simple Self-Dual model, which provide clear testing grounds for the role of the thermodynamic and the continuum limits in the use of the localization measures $z_x$ and $z_t$. We believe that the simplicity and clear localization properties of these models make them useful pedagogical resources for the study of localization transitions.

As discussed in the main text, the values of $z_t$ for single-particle eigenstates of a system may be used to infer the localization properties of the many-body ground state of models whose single-particle eigenstates are mostly localized in momentum space, such as the DDM. Additionally, even in models where the eigenstates are mostly localized in position space, momentum-based quantities such as $z_t$ may provide further details about the momentum distribution. Therefore our result provides a simple way to achieve a more thorough characterization of localization. In particular, one might use $z_t$ to gain useful intuition about the localization properties of nonlocal models which lack a simple position-space description.

Finally, it is important to acknowledge that our discussions are restricted to one-dimensional systems with periodic boundary conditions. In higher dimensional systems or in systems with open boundaries, the correspondence between localization and entanglement properties is not necessarily one-to-one. While we still expect generic delocalized states, having $z_t\to1$, to be LRE, not all LRE states are delocalized. For example, Majorana edge modes are localized states at the edges of a one-dimensional chain, but these are entangled with modes at the other edge of the state, making the overall state LRE while the bulk is localized. It is an interesting direction to study to what extent the expectation value of translation may still capture localization and/or entanglement properties of higher dimensional systems or system with open periodic boundary conditions. Other interesting directions include investigating if this framework can be extended to systems that spontaneously break translation symmetry, and for systems where LRE coincides with topological order. In particular, it would be illuminating to precisely characterize the relationship between localization, entanglement, and the momentum distribution in those systems.

\section{Acknowledgments}
TLH thanks Lei Gioia Yang for useful discussions. TLH and AGL thank Multidisciplinary University Research Initiative (MURI) grant N00014-20-1-2325 for support. AGL thanks the AAUW International Fellowship 2025-26 for support.

\bibliography{momentum}

@misc{KwantFirstSteps,
  author       = {{Kwant Project}},
  title        = {First Steps},
  howpublished = {\url{https://kwant-project.org/doc/1/tutorial/first_steps}},
  year         = {2023},
  note         = {Kwant documentation}
}

@book{Fradkin:2021zbi,
    author = "Fradkin, Eduardo",
    title = "{Quantum Field Theory: An Integrated Approach}",
    isbn = "978-0-691-14908-0",
    publisher = "Princeton University Press",
    month = "3",
    year = "2021"
}

@book{Coleman_2015, place={Cambridge}, title={Introduction to Many-Body Physics}, publisher={Cambridge University Press}, author={Coleman, Piers}, year={2015}}

@misc{tong2016lecturesquantumhalleffect,
      title={Lectures on the Quantum Hall Effect}, 
      author={David Tong},
      year={2016},
      eprint={1606.06687},
      archivePrefix={arXiv},
      primaryClass={hep-th},
      url={https://arxiv.org/abs/1606.06687}, 
}

@article{Aharonov:1969qx,
    author = "Aharonov, Y. and Petersen, A. and Pendleton, H.",
    title = "{Modular variables in quantum theory}",
    doi = "10.1007/BF00670008",
    journal = "Int. J. Theor. Phys.",
    volume = "2",
    pages = "213--230",
    year = "1969"
}

@article{schreiber_exp_bichromatic_2015,
author = {Michael Schreiber  and Sean S. Hodgman  and Pranjal Bordia  and Henrik P. Lüschen  and Mark H. Fischer  and Ronen Vosk  and Ehud Altman  and Ulrich Schneider  and Immanuel Bloch },
title = {Observation of many-body localization of interacting fermions in a quasirandom optical lattice},
journal = {Science},
volume = {349},
number = {6250},
pages = {842-845},
year = {2015},
doi = {10.1126/science.aaa7432},
URL = {https://www.science.org/doi/abs/10.1126/science.aaa7432},
eprint = {https://www.science.org/doi/pdf/10.1126/science.aaa7432}}

@article{goncalves_comensurate_2022,
	title={{Hidden dualities in 1D quasiperiodic lattice models}},
	author={Miguel Gonçalves and Bruno Amorim and Eduardo V. Castro and Pedro Ribeiro},
	journal={SciPost Phys.},
	volume={13},
	pages={046},
	year={2022},
	publisher={SciPost},
	doi={10.21468/SciPostPhys.13.3.046},
	url={https://scipost.org/10.21468/SciPostPhys.13.3.046},
}

@article{philip_RDM_1990,
  title = {Absence of localization in a random-dimer model},
  author = {Dunlap, David H. and Wu, H-L. and Phillips, Philip W.},
  journal = {Phys. Rev. Lett.},
  volume = {65},
  issue = {1},
  pages = {88--91},
  numpages = {0},
  year = {1990},
  month = {Jul},
  publisher = {American Physical Society},
  doi = {10.1103/PhysRevLett.65.88},
  url = {https://link.aps.org/doi/10.1103/PhysRevLett.65.88}
}

@article{comp_chem_overlaps_2016,
author = {Plasser, Felix and Ruckenbauer, Matthias and Mai, Sebastian and Oppel, Markus and Marquetand, Philipp and González, Leticia},
title = {Efficient and Flexible Computation of Many-Electron Wave Function Overlaps},
journal = {Journal of Chemical Theory and Computation},
volume = {12},
number = {3},
pages = {1207-1219},
year = {2016},
doi = {10.1021/acs.jctc.5b01148},
    note ={PMID: 26854874},

URL = {  
        https://doi.org/10.1021/acs.jctc.5b01148
},
eprint = {     
        https://doi.org/10.1021/acs.jctc.5b01148
}

}

@article{boers_bichromatic_2007,
  title = {Mobility edges in bichromatic optical lattices},
  author = {Boers, Dave J. and Goedeke, Benjamin and Hinrichs, Dennis and Holthaus, Martin},
  journal = {Phys. Rev. A},
  volume = {75},
  issue = {6},
  pages = {063404},
  numpages = {6},
  year = {2007},
  month = {Jun},
  publisher = {American Physical Society},
  doi = {10.1103/PhysRevA.75.063404},
  url = {https://link.aps.org/doi/10.1103/PhysRevA.75.063404}
}

@article{harper_1955,
doi = {10.1088/0370-1298/68/10/304},
url = {https://doi.org/10.1088/0370-1298/68/10/304},
year = {1955},
month = {oct},
publisher = {},
volume = {68},
number = {10},
pages = {874},
author = {P G Harper},
title = {Single Band Motion of Conduction Electrons in a Uniform Magnetic Field},
journal = {Proceedings of the Physical Society. Section A}
}

@article{aubry_andre_1980,
author = {Aubry, Serge and André, Gilles},
year = {1980},
month = {01},
pages = {},
title = {Analyticity breaking and Anderson localization in incommensurate lattices},
volume = {3},
journal = {Proceedings, VIII International Colloquium on Group-Theoretical Methods in Physics}
}

@article{sarma_bichromatic_2017,
  title = {Mobility edges in one-dimensional bichromatic incommensurate potentials},
  author = {Li, Xiao and Li, Xiaopeng and Das Sarma, S.},
  journal = {Phys. Rev. B},
  volume = {96},
  issue = {8},
  pages = {085119},
  numpages = {18},
  year = {2017},
  month = {Aug},
  publisher = {American Physical Society},
  doi = {10.1103/PhysRevB.96.085119},
  url = {https://link.aps.org/doi/10.1103/PhysRevB.96.085119}
}

@article{Resta-99,
  title = {Electron Localization in the Insulating State},
  author = {Resta, Raffaele and Sorella, Sandro},
  journal = {Phys. Rev. Lett.},
  volume = {82},
  issue = {2},
  pages = {370--373},
  numpages = {0},
  year = {1999},
  month = {Jan},
  publisher = {American Physical Society},
  doi = {10.1103/PhysRevLett.82.370},
  url = {https://link.aps.org/doi/10.1103/PhysRevLett.82.370}
}

@article{Resta-98-operator,
  title = {Quantum-Mechanical Position Operator in Extended Systems},
  author = {Resta, Raffaele},
  journal = {Phys. Rev. Lett.},
  volume = {80},
  issue = {9},
  pages = {1800--1803},
  numpages = {0},
  year = {1998},
  month = {Mar},
  publisher = {American Physical Society},
  doi = {10.1103/PhysRevLett.80.1800},
  url = {https://link.aps.org/doi/10.1103/PhysRevLett.80.1800}
}

@article{taylor-dimer-momentum-spectrum,
  title = {Characterizing Disordered Fermion Systems Using the Momentum-Space Entanglement Spectrum},
  author = {Mondragon-Shem, Ian and Khan, Mayukh and Hughes, Taylor L.},
  journal = {Phys. Rev. Lett.},
  volume = {110},
  issue = {4},
  pages = {046806},
  numpages = {5},
  year = {2013},
  month = {Jan},
  publisher = {American Physical Society},
  doi = {10.1103/PhysRevLett.110.046806},
  url = {https://link.aps.org/doi/10.1103/PhysRevLett.110.046806}
}

@article{Souza_2000,
   title={Polarization and localization in insulators: Generating function approach},
   volume={62},
   ISSN={1095-3795},
   url={http://dx.doi.org/10.1103/PhysRevB.62.1666},
   DOI={10.1103/physrevb.62.1666},
   number={3},
   journal={Physical Review B},
   publisher={American Physical Society (APS)},
   author={Souza, Ivo and Wilkens, Tim and Martin, Richard M.},
   year={2000},
   month=jul, pages={1666–1683} }

@article{hobbyhorse,
doi = {10.1088/1361-6404/ab1670},
url = {https://dx.doi.org/10.1088/1361-6404/ab1670},
year = {2019},
month = {jun},
publisher = {IOP Publishing},
volume = {40},
number = {4},
pages = {045403},
author = {G A Domínguez-Castro and R Paredes},
title = {The Aubry–André model as a hobbyhorse for understanding the localization phenomenon},
journal = {European Journal of Physics}}

@article{hetenyi-generating_reference,
  title = {Scaling of the bulk polarization in extended and localized phases of a quasiperiodic model},
  author = {Het\'enyi, Bal\'azs},
  journal = {Phys. Rev. B},
  volume = {110},
  issue = {12},
  pages = {125124},
  numpages = {8},
  year = {2024},
  month = {Sep},
  publisher = {American Physical Society},
  doi = {10.1103/PhysRevB.110.125124},
  url = {https://link.aps.org/doi/10.1103/PhysRevB.110.125124}
}

@article{hetenyi-geometric-cumulants,
  title = {Geometric cumulants associated with adiabatic cycles crossing degeneracy points: Application to finite size scaling of metal-insulator transitions in crystalline electronic systems},
  author = {Het\'enyi, Bal\'azs and Cengiz, Serta\ifmmode \mbox{\c{c}}\else \c{c}\fi{}},
  journal = {Phys. Rev. B},
  volume = {106},
  issue = {19},
  pages = {195151},
  numpages = {13},
  year = {2022},
  month = {Nov},
  publisher = {American Physical Society},
  doi = {10.1103/PhysRevB.106.195151},
  url = {https://link.aps.org/doi/10.1103/PhysRevB.106.195151}
}

@article{nonzero_momentum,
  title = {Nonzero Momentum Requires Long-Range Entanglement},
  author = {Gioia, Lei and Wang, Chong},
  journal = {Phys. Rev. X},
  volume = {12},
  issue = {3},
  pages = {031007},
  numpages = {13},
  year = {2022},
  month = {Jul},
  publisher = {American Physical Society},
  doi = {10.1103/PhysRevX.12.031007},
  url = {https://link.aps.org/doi/10.1103/PhysRevX.12.031007}
}

@article{2008-Massar-Spindel,
  title = {Uncertainty Relation for the Discrete Fourier Transform},
  author = {Massar, Serge and Spindel, Philippe},
  journal = {Phys. Rev. Lett.},
  volume = {100},
  issue = {19},
  pages = {190401},
  numpages = {4},
  year = {2008},
  month = {May},
  publisher = {American Physical Society},
  doi = {10.1103/PhysRevLett.100.190401},
  url = {https://link.aps.org/doi/10.1103/PhysRevLett.100.190401}
}

@article{clock_shift_2017,
doi = {10.1088/1674-1056/26/11/117201},
url = {https://dx.doi.org/10.1088/1674-1056/26/11/117201},
year = {2017},
month = {oct},
publisher = {Chinese Physical Society and IOP Publishing Ltd},
volume = {26},
number = {11},
pages = {117201},
author = {Gong, Long-Yan and Ding, You-Gen and Deng, Yong-Qiang},
title = {Uncertainties of clock and shift operators for an electron in one-dimensional nonuniform lattice systems*},
journal = {Chinese Physics B}
}

@article{Concurrence-disordered-2008,
author = {Varga, Imre and Méndez-Bermúdez, José Antonio},
title = {Entanglement in disordered systems at criticality},
journal = {physica status solidi c},
volume = {5},
number = {3},
pages = {867-870},
year = {2008},
doi = {https://doi.org/10.1002/pssc.200777589},
url = {https://onlinelibrary.wiley.com/doi/abs/10.1002/pssc.200777589},
eprint = {https://onlinelibrary.wiley.com/doi/pdf/10.1002/pssc.200777589},

}

@article{momentum-space-shannon-entropy-2012,
  title = {Comparison of Shannon information entropies in position and momentum space for an electron in one-dimensional nonuniform systems},
  author = {Gong, Longyan and Wei, Ling and Zhao, Shengmei and Cheng, Weiwen},
  journal = {Phys. Rev. E},
  volume = {86},
  issue = {6},
  pages = {061122},
  numpages = {9},
  year = {2012},
  month = {Dec},
  publisher = {American Physical Society},
  doi = {10.1103/PhysRevE.86.061122},
  url = {https://link.aps.org/doi/10.1103/PhysRevE.86.061122}
}

@article{PRA_uncertainty,
  title = {Uncertainty relations for general unitary operators},
  author = {Bagchi, Shrobona and Pati, Arun Kumar},
  journal = {Phys. Rev. A},
  volume = {94},
  issue = {4},
  pages = {042104},
  numpages = {11},
  year = {2016},
  month = {Oct},
  publisher = {American Physical Society},
  doi = {10.1103/PhysRevA.94.042104},
  url = {https://link.aps.org/doi/10.1103/PhysRevA.94.042104}
}

@misc{chen_theory_2025,
    title = {Theory of {Localized} {States} in {Quasiperiodic} {Lattices}},
    url = {http://arxiv.org/abs/2509.05950},
    doi = {10.48550/arXiv.2509.05950},
    urldate = {2025-09-09},
    publisher = {arXiv},
    author = {Chen, Jin-Rong and Guo, Xin-Yu and Ding, Shi-Ping and Wu, Tian-Le and Liang, Miao and Gao, Jin-Hua and Xie, X. C.},
    month = sep,
    year = {2025},
    note = {arXiv:2509.05950 [cond-mat]},
}

\newpage
\onecolumngrid

\appendix

\section{Uncertainty relation for unitary operators} \label{appx: uncertainty relation}

Here we review the uncertainty relations for the operators $\hat{U}=e^{i\frac{2\pi}{L}\hat{X}}$ and $\hat{T}=e^{ia\hat{K}}$, using the results for unitary operators and modular variables in Refs. \cite{2008-Massar-Spindel,PRA_uncertainty}. The uncertainties associated with those operators are defined as
\begin{align}
    \Delta U^2 &= \langle \psi|U^\dagger U|\psi\rangle - \langle \psi|U^\dagger|\psi\rangle\langle\psi|U|\psi\rangle\\
    &= 1 - |\langle\psi|U|\psi\rangle|^2 \\
    &= 1-z_x^2,\\
    \Delta T^2&=1-z_t^2,
\end{align}
where $|\psi\rangle$ is some generic many-body state. Note that these uncertainties are not the second cumulants (or ``spread") of the position and momentum distributions themselves. As discussed in Section \ref{sec: prob distr and char funcs}, $z_x$ and $z_t$ can only be used to recover $C_2^{(X)}$ and $C_2^{(K)}$ exactly, as in Eqs. \ref{eq: Resta z} and \ref{eq: c2 momentum simple}, in the thermodynamic and continuum limits, respectively. Indeed, all the moments or all the cumulants, obtained from the characteristic functions as in Eqs. \ref{eq: moment continuous} and \ref{eq: cumulant continuous}, are generally necessary to recover the full probability distributions \cite{Souza_2000, hetenyi-generating_reference}, and $\langle U\rangle = f(k=\frac{2\pi}{L})$ and $\langle T\rangle = f(x=a)$ are only particular values of the characteristic functions of the position and momentum distributions, respectively. Nevertheless, the uncertainty relation between $U$ and $T$ provides an inequality relating $z_x$ and $z_t$ that illuminate the allowed values of those quantities relative to each other, as we explain next.

The operators $U$ and $T$ obey the commutation relation 
\begin{equation}
    UT = e^{-i2\pi\frac{a}{L}\hat{N}}TU,
\end{equation}
where $\hat{N}_e$ is the total charge operator. Using the property that $[\hat{N},U]=[\hat{N},T]=0$ and assuming that the state $|\psi\rangle$ is $U(1)$-symmetric with a definite total charge $N_e$, we can replace the operator $\hat{N}$ with the total charge $N_e$. Then whenever 
$\frac{a{N}_e}{L}\notin \mathbb{Z}$, $U$ and $T$ do not commute. If the state is translation invariant and $\frac{a{N}_e}{L}\notin \mathbb{Z}$, the filling is incommensurate, meaning that there is fractional charge in each unit cell. This implies that the state is long-range entangled \cite{nonzero_momentum}. Defining
\begin{align}
    &\Phi \equiv 2\pi\frac{aN_e}{L}\mod2\pi, \ \ \ \ \ \ \ \ \ 0\leq \Phi< 2\pi,\\
    &A\equiv \left|\tan\frac{\Phi}{2}\right|, \ \ \ \ \ \ \ \ \ \ \ \ \ \ \ \ \ \ \ \ 0\leq A\leq\infty,
\end{align}
we can write the uncertainty relation derived in Refs. \cite{2008-Massar-Spindel, PRA_uncertainty} as
\begin{align}
    &(1+2A)\Delta U^2\Delta T^2 + A^2(\Delta U^2 + \Delta T^2)\geq A^2, \label{eq: uncertainty relation 1}\\
    &(1+2A)(1+z_x^2z_t^2)-(1+A)^2(z_x^2+z_t^2)+A^2\geq 0.
\end{align}

We can now see what this implies for the relationship between $z_x$ and $z_t$. First, suppose that $\Phi=0$, which implies that $[U,T]=0$, and $A=0$. Then the relation
\begin{equation}
    \Delta U^2\Delta T^2 \geq 0,
\end{equation}
imposes no restriction on the possible values of $z_x$ and $z_t$, since by definition those are already $0\leq z_{x,t}\leq 1$. This means that the uncertainties $\Delta U$, and $\Delta T$ can vanish simultaneously. Additionally, if the thermodynamic and continuum limits are concurrently taken, such that Eqs. \ref{eq: Resta z} and \ref{eq: c2 momentum simple} exactly relate $z_{x,t}$ to the spread of their respective distributions, this implies that the spread of the periodic position and momentum distributions can both be narrowly peaked ($z_x\to1$ and $z_t\to1$) at the same time. This is the case for translation symmetric systems with completely filled bands, in which case $z_t=1$ from the translation symmetry, and $z_x=1$ from the commensurate filling, indicating localization in position space, while also having a delta function momentum distribution. Thus, the intuition from the standard position and momentum uncertainty principle, in which a narrow position distribution necessarily implies a wide momentum distribution, is not precise for periodic systems where $[U,T]=0$.

The situation changes when $U$ and $T$ do not commute. Consider a state that is localized, with $z_x=1$ in the thermodynamic limit. Then $\Delta U^2=0$ and Eq. \ref{eq: uncertainty relation 1} becomes
\begin{align}
    A^2\Delta T^2&\geq A^2,\\
    \Delta T^2&\geq 1,
\end{align}
where we used that $A\neq0$ for $[U,T]\neq0$. The last inequality implies that $\Delta T^2 =  1$ and $z_t=0$, since $\Delta T^2$ is bounded above by $1$. Therefore, a state that is completely localized in position while $aN_e/L\notin \mathbb{Z}$ necessarily has a flat many-body momentum distribution. Similarly, if $z_t=1$ while $[U,T]\neq0$, then $z_x=0$. That is, a state with a vanishing momentum second cumulant $C^{(K)}_2$ is necessarily delocalized in position space when the filling is $aN_e/L\notin \mathbb{Z}$. Hence, the intuition from the Heisenberg uncertainty principle can be used to heuristically explain the relationship between $z_x$ and $z_t$ when the operators $U$ and $T$ do not commute.

Finally, we can ask what happens when $U$ and $T$ \textit{almost} commute, in the sense that $\Phi\ll1$. Assuming the usual constraint that $a\ll L$, this situation arises for single-particle states, since $aN_e/L = a/L\ll1$ in that case. Since $A\approx|\Phi/2|$ for $\Phi\ll1$, we can write the uncertainty relation in Eq. \ref{eq: uncertainty relation 1} to leading order in $A$ as
\begin{align}
    &(1+2A)\Delta U^2\Delta V^2\geq0,\\
    &\Delta U^2\Delta V^2\geq 0,\label{eq: uncertainty first order}
\end{align}
where we used the fact that $(1+2A)>0$. Since both uncertainties are always nonnegative, this places no constraints in the spread of the modular momentum or position distributions. Consequently, $z_x$ and $z_t$ can be simultaneously close to $1$. This is the case for the single-particle eigenstates of the DDM at the lower edge of the spectrum, as shown in Fig. \ref{fig: DDM single particle} and discussed in Section \ref{subsec: DDM}. Note that this does not allow both distributions to be delta functions at the same time, which would require $[U,T]=0$ and Eq. \ref{eq: uncertainty first order} to hold exactly at any order in $A$.

\section{Localization of single-particle states} \label{appx: single-particle localization}

\begin{figure*}
    \centering
    \includegraphics[width=0.5 \linewidth]{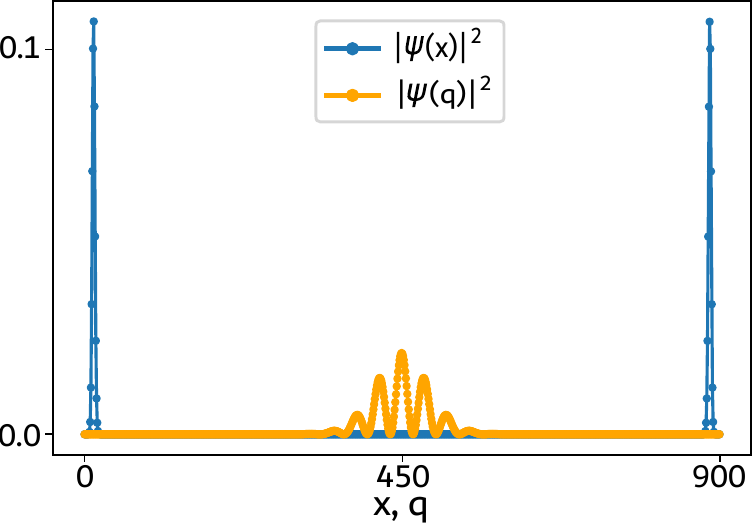}
    \caption{ Position (blue) and momentum (orange) distribution of the lowest-energy  single-particle state of the DDM at $V/t=0.02$. Both the position and the momentum distributions have nonzero values in narrow regions of position or momentum space, making both $z^{sp}_x$ and $z^{sp}_t$ approximately $1$. }
    \label{fig: sp distr 0}
\end{figure*}

Note that Figs. \ref{fig: DDM single particle}a and \ref{fig: DDM single particle}b show the eigenstates at the lowest and highest energies as having both $z_x\to1 $ and $z_t\to1$. The position and momentum probability distributions of the lowest-energy eigenstate of the DDM at $V/t=0.02$, a representative of this phenomenon, are shown in Fig. \ref{fig: sp distr 0}. The figure shows that, while neither the position nor the momentum probability distributions are delta functions, both distributions have peaks concentrated in a small region of the periodic space where they are defined. This feature of the distributions is consistent with both $\lambda_x/L\to 0$ and $\lambda_k/(2\pi/a)\to0$ at the edges of the spectrum, as shown in Fig. \ref{fig: DDM single particle}c.

The simultaneous localization in position and momentum space of the single-particle states is allowed by the uncertainty relation. Indeed, Eq. \ref{eq: UT commutation} implies that, for a single-particle state with $N_e=1$, we have
\begin{align}
    UT = e^{-i2\pi\frac{a}{L}}TU,
\end{align}
which is a trivial commutation relation in the limit $a/L\to0$. Since $U$ and $T$ (almost) commute, the uncertainty relation between those operators places only loose constraints on the allowed values of $z_x$ and $z_t$, as detailed in Appendix \ref{appx: uncertainty relation}. Therefore, $z_x$ and $z_t$ can be simultaneously close to one, meaning that the position and momentum distributions can be localized in position and momentum space. Note that this does not allow both distributions to be delta functions at the same time, which would require $[U,T]=0$ exactly.

\section{Perturbation theory for the DDM}\label{appx: details of PT DDM}
Here we expand on how to predict the properties of the DDM from perturbation theory. As discussed in Section \ref{subsec: DDM single-particle} of the main text, only the unperturbed eigenstates having $|k|>\Delta/2$ or $|k-\pi|>\Delta/2$ do not have their degeneracy lifted at first order, while the other states do. We explicitly show how the $z_t$ plots can be reconstructed with this information alone in the perturbative regime $V/t\ll1$. 

\begin{figure}
    \centering
    \includegraphics[width=\linewidth]{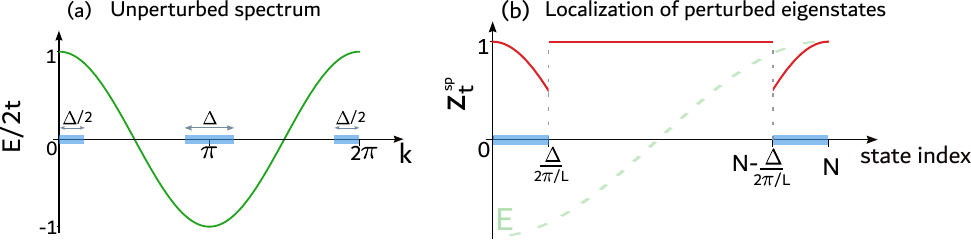}
    \caption{(a) Spectrum of the unperturbed Hamiltonian $H_0=\sum_k2t\cos(ka)c^\dagger_kc_k$. The shaded blue regions correspond to the values of momentum $k$ for which the plane waves have their degeneracy lifted once the perturbation is turned on. These are the states that assume the form $|\tilde{\psi}_{|k|}\rangle$ of Eq. \ref{eq: superposition} once the momentum-coupling term is turned on in the DDM. The parameter $\Delta$ determines the number of other plane waves that couple to the unperturbed plane wave $|\phi_k\rangle$, and which plane waves get their degeneracy lifted. (b) Single-particle $z_t^{sp}$ for the eigenstates of the DDM when the momentum coupling term is turned on. The states marked by the blue regions on the x-axis correspond to those in figure (a), which have the form $|\tilde\psi_{|k|}\rangle$. As discussed in the main text, those states have $z_t^{sp}=|\cos(ka)|$ as discussed around Eq. \ref{eq: zt v function}. The remaining states, which did not have their degeneracy lifted, remain approximate, only slightly widened plane waves of the form $|\psi_{(k)}$, having $z_t^{sp}=1$.}
    \label{fig:schematic PT DDM}
\end{figure}

\begin{figure}
    \centering
    \includegraphics[width=\linewidth]{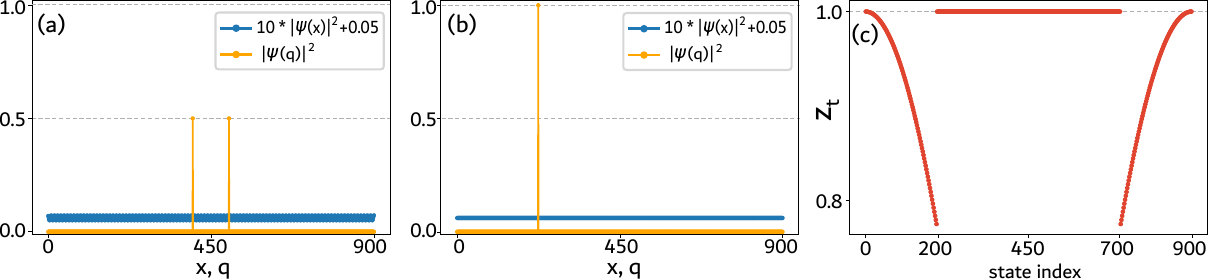}
    \caption{ Position and momentum distributions and the $z_t^{sp}$ of the single-particle eigenstates of the DDM in the perturbative regime, with $V/t=0.00001$, $\Delta=200\frac{2\pi}{L}$, and system size $L=900a$.  
    (a) Position and momentum distributions of the $100^\text{th}$ single-particle eigenstate. This eigenstate is of the type $|\tilde\psi_{|k|}\rangle$, and the momentum distribution is an equal-weight superposition of plane waves with opposite momenta. The peaks are not visibly smeared by the perturbation, because at such low values of $V/t$, the weight of the smearing is negligible.(b) Position and momentum distributions of the $450^\text{th}$ single-particle eigenstate. This is an eigenstate of the type $|\psi_{(k)}\rangle$, which remains approximately a plane wave in the perturbative regime. Indeed, the momentum distribution is simply a delta function in momentum space, as shown by the orange curve. (c) Localization property of all eigenstates of the DDM, in order of increasing energy. The number of eigenstates that have the form $|\tilde \psi_{|k|}\rangle$, like in (a), is exactly $\Delta/(2\pi/L)=200$ at the lower and upper ends of the energy spectrum, in agreement with the expectation from perturbation theory shown in the schematic drawing in Fig. \ref{fig:schematic PT DDM}c. The remaining states are approximate plane waves like in figure (b), and thus have $z_t^{sp}=1$, as expected.}
    \label{fig: DDM perfect perturbation theory}
\end{figure}

In Fig. \ref{fig:schematic PT DDM}, we see the spectrum of the unperturbed Hamiltonian, and the location of the states that have their degeneracy lifted are marked by the blue dashes on the momentum axis. Note that these states correspond to the lowest and highest energy eigenstates of the unperturbed Hamiltonian. When the momentum coupling perturbation $V$ is turned on, the states marked by the blue dash become equal-weight superpositions of the form $|\tilde\psi_{|k|}\rangle$ (Eq. \ref{eq: superposition}) of the approximate plane waves with opposite momenta, as shown in Fig. \ref{fig: DDM perfect perturbation theory}a.
The remaining eigenstates not marked in blue, remain approximate plane waves, as discussed in Section \ref{subsec: DDM single-particle}. The position and momentum distributions of one such state is shown in Fig. \ref{fig: DDM perfect perturbation theory}b, which makes it clear that these states are practically plane waves even after the perturbation is turned on while $V/t\ll1$.
In those figures, $V/t= 0.0001$, well in the perturbative regime, unlike the distributions in Fig. \ref{fig: DDM single patcl distr} of the main text, where $V/t=0.021$. In the latter, the peaks are more widely spread and the superpositions are unequal away from the perturbative regime.

The localization properties expected of the perturbed eigenstates are shown in Fig. \ref{fig:schematic PT DDM}b. The eigenstates that become equal superpositions of opposite plane waves are still marked with blue dashes on the $x$-axis, where the states are ordered by increasing eigen-energies. For these states of the form $\frac{1}{\sqrt{2}}\left(\ket{\phi_k}+\ket{\phi_{-k}}\right)$, we expect $z_t=|\cos ka|$ as shown in Fig. \ref{fig:schematic PT DDM}b. We verify that this expectation is correct by plotting $z_t$ for $V/t=0.00001$ and $\Delta=200\frac{2\pi}{L}$ in Fig. \ref{fig: DDM perfect perturbation theory}c. Note that there are exactly $200$ eigenstates at the edges of the spectrum that follow the $z_t=|\cos ka|$ curve, which matches the expectation from Fig. \ref{fig:schematic PT DDM}b that those correspond to the equal superposition eigenstates. Additionally, we see that all the approximate plane waves in the middle of the spectrum indeed have $z_t=1$.

\begin{figure}
    \centering
    \includegraphics[width=0.5\linewidth]{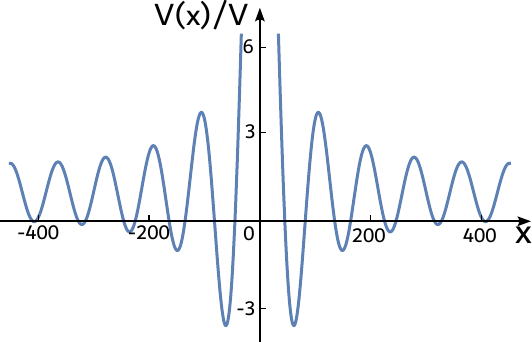}
    \caption{Onsite potential in position space $V_\text{pos}(x)$ of the DDM, in units of the momentum coupling term coefficient $V$. Note that the potential is not symmetric around the $x$-axis. Instead, the oscillations are shifted upwards, which skews the spectrum of the DDM to having higher energies. In the figure, we used $\Delta=10/(2\pi/L)$ and $L=900$. For larger $\Delta$ at fixed system size, the oscillations increase in frequency.}
    \label{fig:onsite potential}
\end{figure}

Finally, we briefly comment on what happens beyond the perturbative regime, which is the case studied in the main text. Note that the $z_t$ curves shown in Fig. \ref{fig: DDM single particle}b are not symmetric about the center of the spectrum, unlike in the perturbation theory regime shown in Fig. \ref{fig: DDM perfect perturbation theory}c. This happens because, at higher values of the parameter $V$, the onsite potential in position space $V_\text{pos}(x) = \sum^\Delta_{\delta=0} 2V\cos(\delta x)$, shown in Fig. \ref{fig:onsite potential}, skews the spectrum to higher values, breaking its time-reversal symmetry. The finite-size effects of this potential explain the asymmetry in the localization of the single-particle states, which are affected by the skewing of the eigen-energies.

\section{Conditions for $z^{MB}_t=1$ in the DDM} \label{appdx: ddm z=1}

Here we show that the many-body ground state of the DDM is delocalized, with $z^{MB}_t\approx 1$, when all the single-particle eigenstates having energies $E\leq E_{V,1}$ are filled, i.e., when $\mu=E_{V,1}$. Recall that
\begin{align}
    z^{MB}_t=|\langle \Psi|\hat{T}|\Psi\rangle|=|\det S|,
\end{align}
for $S_{ij}=\langle\psi_i|\hat{T}|\psi_j\rangle$ and $\{|\psi_n\rangle\}$ the set of the occupied single-particle eigenstates in the ground state. So we need to show that $|\det S|\approx1$.

Note that $S$ is a submatrix of the unitary matrix representing the translation operator $T$, written in the eigenbasis of the DDM. This submatrix is not itself unitary unless $\langle\psi_i|T|\psi'_{l}\rangle=0$, where $|\psi_i\rangle$ is any occupied eigenstate and $|\psi'_l\rangle$, any unoccupied eigenstate. That is, unless $S$ is a diagonal block of the unitary matrix representation of the translation operator $T$, written in the eigenbasis of the DDM.  The condition that $S$ is unitary can be expressed as the condition that the occupied DDM single-particle eigenstates involve superpositions of only plane waves which do not also appear in the superpositions of the unfilled DDM eigenstates. To see this, let $\bar{W}$ be the matrix that diagonalizes the complete translation matrix $T$, written in the DDM eigenbasis, into the eigenbasis of translation. When $S$ is a diagonal block of $T$ in the eigenbasis of the DDM, the matrix $\bar{W}$ also has a diagonal block $W$ which diagonalizes $S$ into the eigenbasis of translation. This is analogous to the matrix $W$ used in Eq. \ref{eq: Sw} in Section \ref{DDM subsec 2}, where we discuss how the uncoupling of the filled single-particle states into plane waves implies $|\det S|=1$. In fact, if $\bar{W}$ does not have such a diagonal block, then the set of filled single-particle DDM eigenstates must involve superpositions of plane waves which also appear in the unfilled states. In that case, $S$ is not a diagonal block of $T$. Conversely, if $\bar W$ is block-diagonal and $W$ is one of its diagonal blocks, then $T$ is block-diagonal with $S$ being one of those blocks, and, consequently, a unitary matrix.

Here, we claim that $S$ is approximately a diagonal block of the unitary $T$ when all the lower-energy eigenstates of the DDM of the form $|\tilde\psi_{|k|}\rangle$, i.e., superpositions of plane waves of opposite momenta, are filled. That is, all eigenstates with energy $E\leq E_{V,1}$ are filled. We claim that $S$ is only \textit{approximately} a diagonal block of $T$, in the sense that there are still some small but non-vanishing terms $\langle\psi_i|T|\psi'_{l}\rangle$, because the filled eigenstates at $\mu=E_{V,1}$ still involve plane waves which appear in the unfilled states. We will make these statements more precise next.

As discussed in Section \ref{subsec: DDM}, the eigenstates at $E<E_{V,1}$ have the form $|\tilde\psi_{|k|}\rangle$. These are superpositions of two sets plane waves $|\psi_{(k)}\rangle$ and $|\psi_{(-k)}\rangle$, with weight near $k$ and $-k$ respectively, as in Eq. \ref{eq: ddm eigenstate psi}, reproduced here for convenience:
\begin{align}
    |\psi_{(k)}\rangle=|\phi_k\rangle&+ \frac{V}{t}\sum_{\delta=\frac{2\pi}{L}}^{\Delta}\big[\frac{1}{\cos(ka)-\cos((k+\delta)a)}|\phi_{k+\delta}\rangle\nonumber\\
   & +\frac{1}{\cos(ka)-\cos((k-\delta)a)}|\phi_{k-\delta}\rangle\big ].
\end{align}
Then when one eigenstate of the form $|\tilde\psi_{|k|}\rangle$ is filled at $E<E_{V,1}$, all other eigenstates which contain the same plane waves $|\phi_{\pm k\pm \delta}\rangle$ in their superposition must also be filled to guarantee $\langle\psi_i|T|\psi_{j\neq i}\rangle=0$. Furthermore, as those eigenstates are filled, they introduce the need of filling new eigenstates that contain plane waves with momenta not included in the previous one.

Each new eigenstate of the form $|\tilde\psi_{|k|}\rangle =\frac{1}{\sqrt{2}}\left(|\psi_{(k)}\rangle+|\psi_{(-k)}\rangle\right)$ that is filled at $E<E_{V,1}$ introduces new pairs of plane waves in the outer edge of its momentum distribution, near momenta $\pm|k+\Delta|$, that were not in the set of previously filled plane waves. In addition to the states $|\tilde\psi_{|k'|}\rangle$ at $k'\neq k$ needed to account for those new plane waves, there is also the need to fill the state $|\tilde\psi'_{|k|}\rangle=\frac{1}{\sqrt{2}}\left(|\psi_{(k)}\rangle-|\psi_{(-k)}\rangle\right)$ which was obtained from the originally degenerate pair of the unperturbed eigenstate $|\phi_k\rangle$, since they encompass the superposition of the same plane waves. Therefore, until all eigenstates at $E<E_{V,1}$ are filled, the set of plane waves $\{|\phi_k^{f}\rangle\}$ that are in the superpositions of the filled states, and the set of plane waves $\{|\phi_k^{{nf}}\rangle\}$ that are in the superpositions of the eigenstates that are not filled have a non-vanishing intersection $\{|\phi_k^{f}\rangle\}\cap\{|\phi_k^{{nf}}\rangle\}\neq\{\}$.

 After $E=E_{V,1}$, the next lowest energy eigenstates have momentum distributions with single smeared peaks, of the form $|\psi_{(k)}\rangle$. We find numerically (and can argue from perturbation theory) that the spread around each peak is the same for eigenstates having one or two peaks in their momentum distribution. Hence,  the set of plane waves involved in the superpositions of states of the type $|\psi_{(k)}\rangle$ is smaller than the set for states of the type $|\tilde \psi_{|k|}\rangle$ at $E<E_{V,1}$. While the states $|\psi_{(k)}\rangle$ still involve plane waves that were not previously filled, those plane waves contribute to the superposition of a smaller number of unfilled states, since their momentum distributions are restricted to smaller ranges of $k$. Then for each eigenstate $|\psi_{(k)}\rangle$ that is filled at $E_{V,1}<E<E_{V,2}$, only a small number of unfilled other eigenstates will have any overlap  with the newly added plane waves of the set $\{\phi^f_k\}$, as the eigenstates in this energy range are approximately plane waves in our parameter range. 

Hence, when all of the eigenstates having $E<E_{V,1}$ and any number of eigenstates having $E_{V,1}<E<E_{V,2}$ are filled, $S$ is approximately a unitary matrix and has $|\det S|\approx1$. Therefore, we expect ground states of the DDM at chemical potential $\mu$ satisfying $E_{V,1}<\mu<E_{V,2}$ to have $z_t^{MB}\approx1$, which corresponds to what is observed in Fig. \ref{fig: DDM phase diagrams}. 

\section{Continuum and thermodynamic limits} \label{appx: limits}
Here we describe how the limits are generically taken for the tight-binding models considered in this work where the limits can be defined. For some models, such as the Deterministic Dimer Model, extra care has to be taken with the momentum hopping term to ensure that the localization transitions happen at the same onsite potential strength as the thermodynamic limit is taken. Here we merely describe the general procedure of taking each limit.

For a generic 1D periodic lattice model, we take the thermodynamic limit by multiplying the original system size $L_0$ by a real constant $c\geq 1$ while keeping the lattice spacing $a_0$ fixed. That is, $L_0=a_0N_0$ becomes $L=cL_0 = a_0N$, for $N=cN_0$ sites in the lattice. In the continuum limit, on the other hand, the system size $L_0$ is left invariant, while $a_0\to a=a_0/c$. This also leads to an increase of sampled points $N$ across the potential lattice, such that $L = aN=\frac{a_0}{c}N=L_0$ implies that $N=cN_0$. That is, $N$ increases by the same factor in either the continuum or thermodynamic limits. Additionally, in the continuum limit, the position hopping coefficient $t_0$ has to be appropriately scaled as $t_0\to t=c^2t_0$. The reason is that $t\propto 1/a^2$, so it scales with the lattice spacing as well. This can be seen from the discretization of the kinetic term of a continuous Hamiltonian in the tight-binding limit:
\begin{align}
    -\frac{\hbar^2}{2m}\partial^2 \to-\frac{\hbar^2}{2m}\frac{1}{a^2} \sum_{i}\left(|i+1\rangle\langle i| + |i\rangle\langle i+1| - 2|i\rangle\langle i|  \right)
\end{align}
in the limit where $a\to 0$ \cite{KwantFirstSteps}. Therefore, the hopping term is given by 
\begin{equation}
    t = \frac{\hbar^2}{2ma^2}.
\end{equation}
Alternatively, one could restrict the analysis to the long wavelength limit and, from the discrete Hamiltonian, and write $\sum_k2t\cos(ka)c^\dagger_kc_k\approx \sum_kta^2k^2c^\dagger_kc_k+\text{constant energy shift}$. Then, decomposing the creation and annihilation operators for momentum space into fields in the continuous position space \cite{Coleman_2015,Fradkin:2021zbi}, we get
\begin{align}
    \sum_kta^2\frac{1}{L}\int dxdx'
\psi^\dagger(x')\frac{\partial^2}{\partial x^2}e^{ik(x-x')}\psi(x)
&=\int dx (ta^2)\psi^\dagger(x)\partial^2_x\psi(x)\\&=\int dx \ t'\psi^\dagger(x)\partial^2_x\psi(x).
\end{align}
This means that hopping terms of the type $t'c^\dagger_xc_{x+a}$ correspond to $2\frac{t'}{a^2}\cos(ka)c^\dagger_kc_k$ in momentum space. Therefore, when we take $a_0\to a_0/c$, we also take $t\to t_0c^2$ for the hopping term $t$ in the diagonal term of the momentum-space Hamiltonian. Taking the thermodynamic limit, on the other hand, does not require adjusting $t_0$, since $a_0$ is fixed and $t_0$ is independent of $L_0$.

\section{$z_x^{sp}$ continuum and thermodynamic limits} \label{appx: zx_sp}
\begin{figure}
    \centering
    \includegraphics[width=0.75\linewidth]{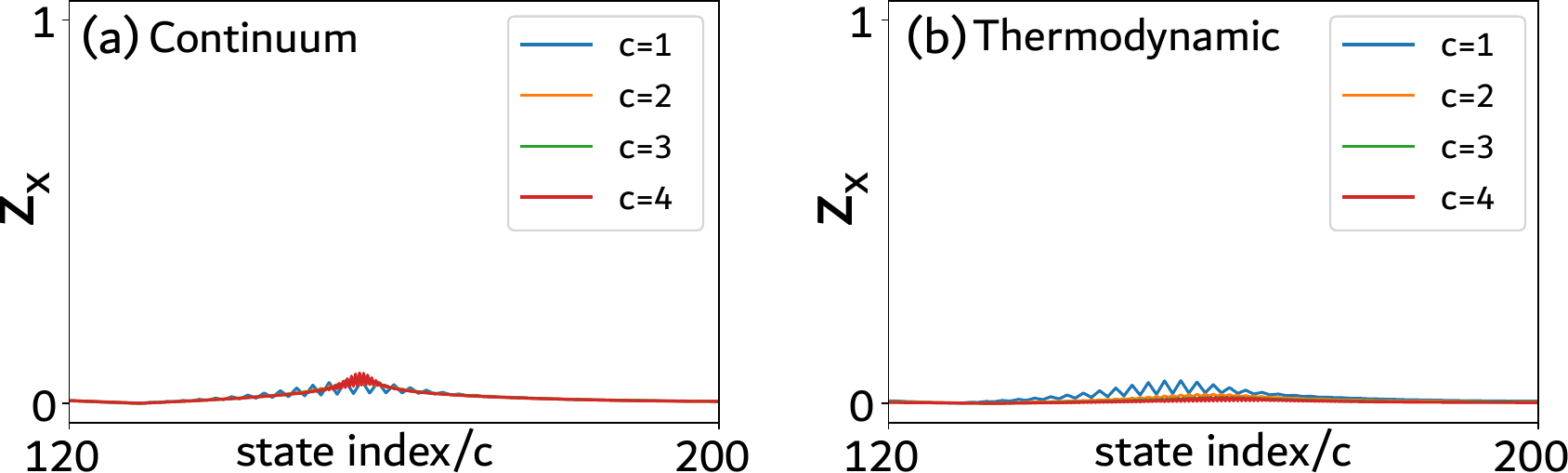}
    \caption{Effects of the continuum (a) and thermodynamic (b) limits on the single-particle $z_x^{sp}$ of the eigenstates of the DDM around energy $E=E_{V,1}$. (a) The continuum limit does not qualitatively change the shape of the $z_x^{sp}$ curve, which remains like that shown in Fig. \ref{fig: DDM single particle}a. The small bump corresponds to the energy $E_{V,1}$, before which the single-particle eigenstates are partially-localized, and after which they are completely delocalized. (b) The thermodynamic limit reduces the amplitude of the bump and shifts the $z_x^{sp}$ curve, such that the small bump accompanies the shift of the single-particle $z_t^{sp}$ shown in Fig. \ref{fig: DDM single particle limits}b in the same limit. }
    \label{fig:app_pair}
\end{figure}

Here we show the single-particle $z_x^{sp}$ of the eigenstates of the DDM in the continuum and thermodynamic limits, for $V_0/t_0=0.071$ (compare with Fig. \ref{fig: DDM single particle limits} for $z_t^{sp}$). As described in the main text, the continuum limit does not qualitatively change the behavior of $z_x^{sp}$, as shown in Fig. \ref{fig:app_pair}a. Indeed, for every $c$ shown in the figure, $z_x^{sp}$ matches the shape of Fig.\ref{fig: DDM single particle}a for the corresponding single-particle states around energy $E_{V,1}$, where the single-particle states go from being partially localized to completely delocalized.

The effect of taking the thermodynamic limit is shown in Fig. \ref{fig:app_pair}b. It might seem counter-intuitive that the thermodynamic limit would lower the values of $z_x^{sp}$ for the single-particle states, since Resta's result Eq. \ref{eq: Resta's formula} implies $z_x\to1$ for localized states in the thermodynamic limit. Recall, however, that the single-particle eigenstates of the DDM are at most partially localized, with a few exceptions at the edges of the spectrum, where $z_x\to1$ in Fig. \ref{fig: DDM single particle}a. Therefore, we do not expect these partially-localized eigenstates to have $z_x^{sp}\to1$. In fact, $z_x\to0$ for those states, as their position distributions are almost uniform, as illustrated in the blue curve of Fig. \ref{fig: DDM single patcl distr}b for a representative partially-localized single-particle state. It is then not surprising that the thermodynamic limit would make $\lambda_x\to\infty$, as the localization length is at least proportional to the system size. The details of why the thermodynamic limit is reducing $z_x^{sp}$ for the specific range of states in Fig. \ref{fig:app_pair}b is beyond the scope of this paper and does not impact our other results. Additionally, the thermodynamic limit shifts the location of the bump, such that it still matches the location of the increase in $z_t^{sp}$, which also shifts in the thermodynamic limit, as shown in Fig. \ref{fig: DDM single particle limits}b.

\end{document}